\documentclass[iop]{emulateapj}

\pdfoutput=1

\slugcomment{To appear in ApJ}

\shorttitle{The low surface brightness frontier}
\shortauthors{Trujillo \& Fliri}

\begin{document}


\title{Beyond 31 mag/arcsec$^2$:  the low
surface brightness frontier with the largest optical telescopes}


\author{Ignacio Trujillo\altaffilmark{1} and J\"uergen Fliri}
\affil{Instituto de Astrof\'{\i}sica de Canarias, c/ V\'{\i}a L\'actea s/n, 
E-38205, La Laguna, Tenerife, Spain}

\affil{Departamento de Astrof\'{\i}sica, Universidad de La Laguna, E-38206, La 
Laguna, Tenerife, Spain}
\email{trujillo@iac.es}




\begin{abstract}

The detection of optical surface brightness structures in the sky with magnitudes fainter than 30 mag/arcsec$^2$
(3$\sigma$ in 10$\times$10 arcsec boxes; r-band) has remained elusive in current photometric deep surveys. Here
we show how present-day 10 meter class telescopes can provide broadband imaging 1.5-2 mag deeper than most
previous results within a reasonable amount of time (i.e. $<$10h on source integration). In particular, we
illustrate the ability of the 10.4 m Gran Telescopio de Canarias (GTC) telescope to produce imaging with a
limiting surface brightness of 31.5 mag/arcsec$^2$ (3$\sigma$ in 10$\times$10 arcsec boxes; r-band) using 8.1
hours on source. We apply this power to explore the stellar halo of the galaxy UGC00180, a galaxy analogous to M31
located at $\sim$150 Mpc, by obtaining a surface brightness radial profile down to $\mu_r$$\sim$33 mag/arcsec$^2$.
This depth is similar to that obtained using star counts techniques of Local Group galaxies, but is achieved at a distance
where this technique is unfeasible. We find that the mass of the stellar halo of this galaxy is
$\sim$4$\times$10$^{9}$M$_{\sun}$, i.e. 3$\pm$1\% of the total stellar mass of the whole system. This amount of
mass in the stellar halo is in agreement with current theoretical expectations for galaxies of this kind.

\end{abstract}





\section{Introduction}

Ongoing technological advances are enabling the observation of deeper data every day,  allowing us to discover objects that
were hidden to previous generations of astronomers. Nowadays, the deepest optical data, the Hubble Ultra Deep Field
\citep[HUDF;][]{2006AJ....132.1729B}, is able to detect point-like objects as faint as 29 mag (10$\sigma$; using 0.2\arcsec
apertures). While our ability to detect compact structures in deep optical surveys is impressive, when the photons spread
over extended areas, the lack of contrast against the foreground sky penalizes our capacity to identify large objects.  

Exploring astronomical objects with low surface brightness is extremely challenging from an observational point
of view. It is not enough to have very deep data but a careful reduction and treatment of the sky is absolutely
necessary. In fact, it is common to find in the literature very deep data where the handling of the sky
(although optimized for the detection of the faintest point-like sources) is inappropriate for the
characterization of the faintest surface brightness structures. In this sense, for instance, it is easy to find
"holes" around the brightest extended galaxies in very deep surveys like the Canada-France-Hawaii Legacy Survey
\citep[CFHTLS;][]{Goranova2009} or the  HST eXtreme Deep Field \citep[][]{2013ApJS..209....6I}. These are
examples where the reduction pipeline has been probably very aggressive on subtracting the sky. This is very
likely due to real signal, coming from low surface brightness features around the objects, being confused with
the background of the image and, consequently, oversubstracted.

The treatment of the sky is not the only actor playing a major role on the ability to detect and characterize the low
surface brightness astronomical structures. In fact, there are many artifacts that affect the quality of the images:
fringing, scattered light, ghosts, etc. All these phenomena generate surface brightness gradients on the images that
enormously complicate the study of the faintest surface brightness components. To surpass all these problems, there have
been an increasingly large number of works addressing these observational difficulties
\citep[e.g.][]{2012ApJS..200....4F,2015MNRAS.446..120D,fliri2016}. All these studies have pointed out the need for a careful
preparation of the observational strategy and the reduction of the data. As the result of these efforts, state-of-the-art
deep surveys aiming to explore the faintest surface brightness structures are currently reaching $\sim$29-30 mag/arcsec$^2$
\citep[3$\sigma$, 10$\times$10 arcsec boxes;
e.g.][]{2010AJ....140..962M,2012ApJS..200....4F,2014ApJ...787L..37M,2015MNRAS.446..120D,2015ApJ...807L...2K,2015A&A...581A..10C}.
Most of these studies have been conducted with telescopes whose apertures range from small ($\sim$1m) to moderate
($\sim$4m). Is it possible to go significantly deeper with present-day largest (i.e. 10m class) telescopes? The goal of the
present work is to explore the depth, in terms of surface brightness, that current largest optical telescopes can achieve
within a reasonable amount of time (i.e. time on source less than 10h).

The study of the stellar halos surrounding nearby galaxies is one of the many reasons to conduct very deep
imaging. Probing the stellar halos in a large number of galaxies is a strong test to the current $\Lambda$CDM
galaxy formation scenario \citep[e.g][]{2005ApJ...635..931B,2006MNRAS.365..747A,2008ApJ...689..936J}. In
fact, state-of-the-art cosmological simulations suggest that virtually all present-day galaxies will show
several streams and a prominent extended stellar halo if they are observed down to $\mu_V$$>$31 mag/arcsec$^2$
\citep[e.g.][]{2010MNRAS.406..744C}. This prediction remains untested except for a very limited number of
galaxies in the Local Group
\citep[][]{2009Natur.461...66M,2009MNRAS.395..126I,2011ApJ...738..150T,2014ApJ...780..128I,2015ApJ...800...13P} where the resolved
star counts technique has been used. In fact, with this technique, these studies have revealed features with
equivalent surface brightness of $\sim$31-32 mag/arcsec$^2$. However, the technique of resolved star counts can
not be applied very far away. Using the HST, \citet{2012MNRAS.421..190Z} have estimated a maximum distance of
16 Mpc for this strategy. This considerably limits the volume, the number and the type of galaxies that can be
studied. For this reason, we need to explore how deep we can go, in terms of surface brightness, with
integrated photometry. In fact, considering the intrinsic stochasticity of the stellar halo formation process,
the need of a larger sample of galaxies is clear if we want to probe the $\Lambda$CDM galaxy formation scenario
in depth.  The current paper is designed as a pilot project to explore whether integrated photometry can
produce images as deep as the star counting technique. We will show that this is indeed feasible, opening the
possibility of exploring the stellar halos to a much larger volume than the Local Group of galaxies. 

This paper is structured as follows. In Section ~\ref{data}, we describe our data, the target selection criteria, the
observational strategy and the data processing. Section  ~\ref{results} shows the results of our observation, and how the
depth of our image compares with other previous surveys. In section \ref{scattered light}, we explore the distribution of the scattered
light in our field of view. The characteristics of the stellar halo of our targeted galaxy, UGC00180,  as well as the effect
of the point spread function (PSF) are described in  Section \ref{stellarhalo}. Section \ref{discussion} discusses the main
results of this paper and, finally, our work is summarized in Sect.\ref{conclusions}. Hereafter, we assume a cosmology with
$\Omega_{\rm m}= 0.3$, $\Omega_{\rm \Lambda}= 0.7$ and H$_0 = 70$ km~s$^{-1}$~Mpc$^{-1}$.

\section{Data}
\label{data}


Ultra deep observations of the galaxy UGC00180 and its surrounded region were carried out with the Gran Telescopio de
Canarias (GTC) using the OSIRIS (Optical System for Imaging and low-Intermediate-Resolution Integrated Spectroscopy) camera.
OSIRIS has a total field of view (FOV)of 7.8\arcmin$\times$8.5\arcmin, of which 7.8\arcmin$\times$7.8\arcmin are
unvignetted. The OSIRIS camera is composed of two CCDs with a gap of 9.4\arcsec between them. The pixel scale of the camera
is 0.254\arcsec. The images were obtained using the Sloan r' filter during 6 (non-consecutive) nights. Images were taken
under good seeing conditions, producing a final image with a Full Width Half Maximum (FWHM) seeing of $\sim$0.9\arcsec. 

\subsection{Target selection}

The selection of the galaxy, UGC00180, was done to assure that the OSIRIS FOV was able to cover a
significant region of sky around the galaxy plus the possibility of exploring very extended stellar
features surrounding the object. UGC00180 is a galaxy positioned at R.A.(2000)= 00h19m08.2s and
Dec.(2000)=+15d44m57.6s. Its redshift is 0.0369. This locates the object at a distance of 151.3 Mpc,
giving a scale of 0.733 kpc/\arcsec. Consequently, a single shot of the OSIRIS camera covers
343$\times$343 kpc at the galaxy distance. In addition, we decided to take this galaxy, with
characteristics similar to the well explored massive galaxies in our vicinity, so we can have a
reference to compare with. According to Hyperleda \citep[][]{2003A&A...412...45P}, UGC00180 is a Sab
massive galaxy (M$_B$=-21.76, V$_{rot}$=267.6$\pm$18.4 km/s). In this sense, this galaxy is comparable
with M31, a massive Sb galaxy (M$_B$=-21.2, V$_{rot}$=256.7$\pm$6.1 km/s). UGC00180 has
R$_{25}$(B-band)=32$\pm$3\arcsec (i.e. 23.5$\pm$2.2 kpc). This, together with its rotational velocity,
translates into a dynamical mass of M$_{dyn}$=(3.9$\pm$0.9)$\times$10$^{11}$ M$_{\sun}$  inside its
optical radius. Its global (Petrosian) color, according to NED, is g-r=0.78 (after correcting by
Galactic extinction). Following \citet{2003ApJS..149..289B}, this color is equivalent to a
(M/L)$_r$=2.51 \citep[Kroupa Initial Mass Function (IMF);][]{2001MNRAS.322..231K}. Consequently, the
stellar mass of UGC00180 is M$_\star$$\sim$1.3$\times$10$^{11}$M$_{\sun}$.

\subsection{Observational strategy}\label{strategy}

The objective of our observation is to reach the theoretical surface brightness limit of GTC within the total
amount of time allocated for this exercise (i.e. $\sim$8.1 hours on source as we describe below). To achieve
this goal, we need to deal with several observational biases that affect very deep observations: fringing,
scattered light, saturation, ghosts, etc. For this reason, we have designed an observational strategy that aims
to obtain a background as flat as possible  around our galaxy target. To do that, in addition to conduct the
usual dithering scheme, we have carried on a rotation pattern to remove as much as possible effects due to
scattered (residual) light contamination.  With the rotation pattern, we avoid that a potential reflected
residual light (from the telescope dome, telescope structure, etc)  affects the camera along the same position
angle (P.A.) during the full set of pointings. Moreover, the large number (243) of images we get in the end
allow us to build a very flat sky to reach our purpose. 

The strategy that we have conducted is as follows. We have carried out 9 observing blocks, each block composed
by three steps.

\begin{itemize}
\item Step A: The P.A. of the camera is fixed to a given angle and we make a dithering pattern of 9 positions.
\item Step B: The P.A. is rotated 120 degrees with respect to the previous angle and we repeat the dithering
pattern of 9 positions.
\item Step C: The P.A. is rotated again another 120 degrees and we repeat a dithering pattern of 9 positions.

\end{itemize}

The offsets of the dithering sequence, both in R.A. as well as in Dec. are of 1\arcmin. This offset is more
than enough to derive a proper background map considering the brightness and size of the astronomical objects
in the field of view of our final image. Moreover, this guarantees that our target of interest, the galaxy
UGC00180, is observed over the 100\% of the time. At the end of each observing block we have 27 images. The
observing blocks are identical to each other but starting with a different set of orientation angles. These are:

\begin{itemize}
\item 0 - 120 - 240
\item 10 - 130 - 250
\item 20 - 140 - 260
\item 30 - 150 - 270
\item 40 - 160 - 280
\item 50 - 170 - 290
\item 60 - 180 - 300
\item 70 - 190 - 310
\item 80 - 200 - 320
\end{itemize}

\begin{figure*}
\includegraphics[width=\textwidth]{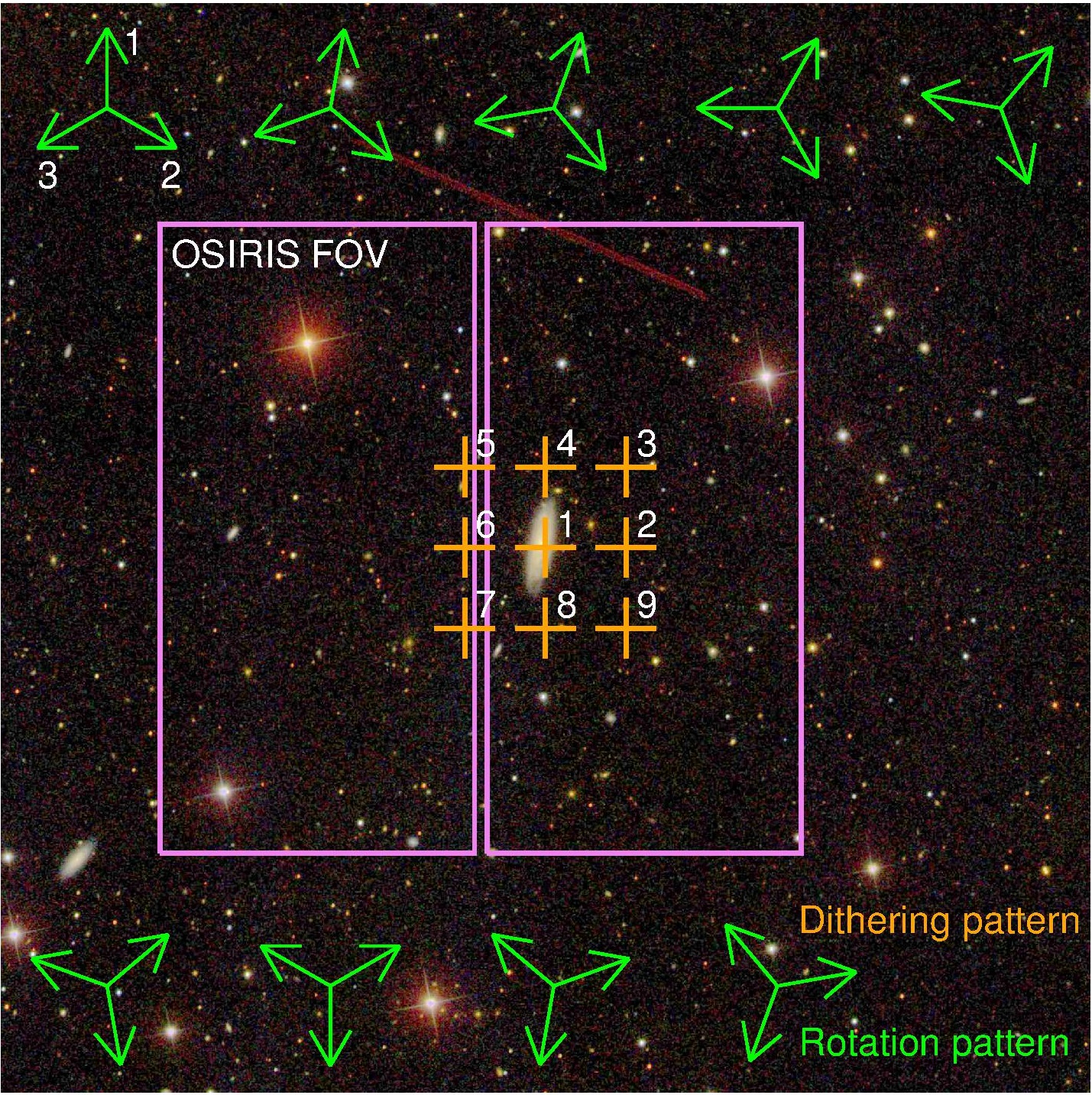}

\caption{Dithering and rotation sequence followed to get the final image. The background image is a color
composite of the UGC00180 region obtained from SDSS. The total field of view of the SDSS image corresponds to
13.5\arcmin$\times$13.5\arcmin. The field of view of each pointing by the OSIRIS camera
(7.8\arcmin$\times$7.8\arcmin) is overplotted with a violet contour. The position of the orange crosses
indicate the dithering pattern followed in each block of observations, whereas the green arrow indicates the
position angle of the camera in each set of observations. In total, the final image is composed by
3$\times$9$\times$9 pointings. A full description of the procedure is done in the text.}

\label{pattern}

\end{figure*}

The observational strategy is illustrated in Fig. \ref{pattern}. Each pointing of the sequence has an
exposure time of 120s.  That corresponds to a total amount of time on source of 8.1 hours. It must be noted
that shorter exposure times than the ones conducted here, although desirable to avoid large sky variation
during each exposure, would prohibitively increase the amount of time allocated to overheads (the camera
readout time is 21 seconds)\footnote{It is worth noting that the optical sky brightness  at the Roque de los
Muchachos observatory is extremely stable during the night. Measured variations are consistent with zero,
0.03$\pm$0.07 mag/arcsec$^2$, within the precision error (see a report at
http://www.ing.iac.es/astronomy/observing/conditions/skybr/skybr.html).}. Consequently, a balance between
both quantities is necessary. In addition to the previous set of images, there were also a number of shorter
exposure times images (6 pointings of 5 seconds each and two pointing of 10 seconds) to avoid saturation of
the central parts of UGC00180.

\begin{figure*}
\includegraphics[width=\textwidth]{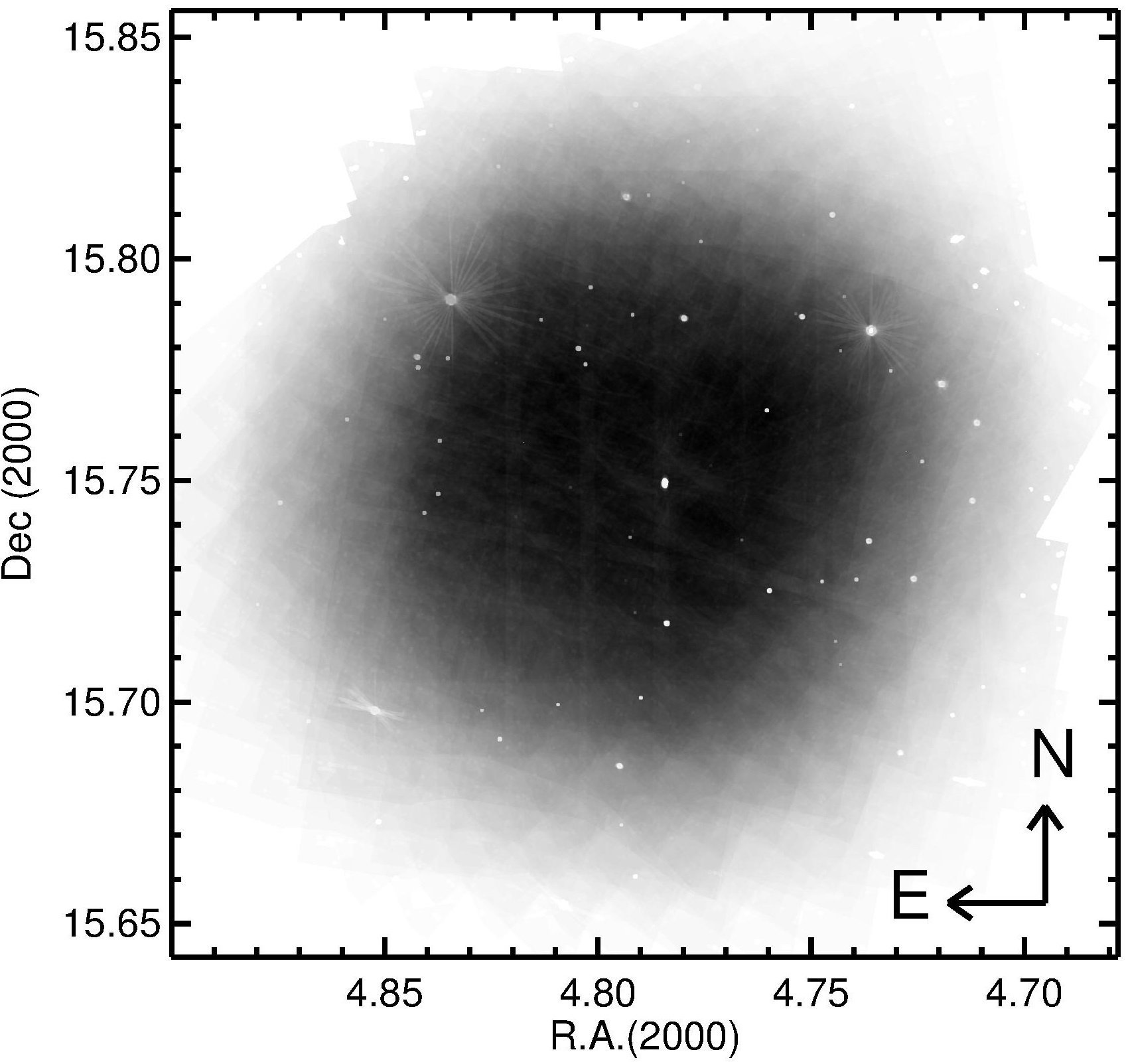}

\caption{Weight map after the stacking process. Darker areas correspond to a larger number of repetitions and,
consequently, deeper imaging. Note the dithering and rotation scheme followed in this observation.}

\label{weightmap}

\end{figure*}

As we have mentioned before, each OSIRIS pointing covers an unvignetted region of
7.8\arcmin$\times$7.8\arcmin. The dithering and rotation sequence that we have conducted makes a final
image covering a larger field of view (12.7\arcmin$\times$12.7\arcmin). Fig. \ref{weightmap} shows the
weight map resulting from the stacking process that we describe in the following sections. Within the
central four arcmin of the image, the amount of observing time per pixel is quite homogeneous, with a
standard deviation per pixel of $\sim$2\%. Our observing pattern allows the galaxy to never occupy
exactly the same physical area of the CCD across the 243 exposures, helping in the building of an
accurate background map. The observation pattern also helps to remove the gaps between the OSIRIS CCD. 

\subsection{Flat-field correction}

An accurate estimation of the flat field correction is key for the purpose of this work. Dome flats are not useful for
our goals due to inhomogeneities in the GTC Dome illumination. Consequently, our flat field correction should be based on
sky imaging. Twilight flats, although better, are still insufficient for our purposes as variations between the
night-sky and the twilight spectrum may result in subtle flat fielding differences. For this reason, we have decided to
use our own set of science images to create a masterflat.

Masterflats are created for each observing night using a median of the normalized science images of that night. Ideally,
one would like to use the full set of science images to obtain a much better masterflat, however, slight differences in
the focus and vignetting correction from night to night prevent such approach. As mentioned before, the total number of
nights used to complete our dataset were 6. The amount of data is such that there are at least  15 science images every
night to combine and create the masterflat of that night. An independent masterflat is created for each CCD of the
OSIRIS camera. The building of the masterflats are done as follows:

\begin{itemize}

\item For each individual 120s science image, we create an object mask using SExtractor \citep{1996A&AS..117..393B}. The
object masks are expanded to assure that the outer light of the objects is also masked. Only those pixels outside the
masks are used to create the final masterflat.

\item Every individual science image of a given night is normalized to one. The determination of the number
counts to normalize each image is done in the same CCD position (close to the optical axis of the camera). This
counting is done within a box of 50$\times$50 arcsec$^2$.

\item The normalized and masked individual science images are combined in a single masterflat using the median.

\end{itemize}

The masterflats have a typical rms of 0.055\%. Finally, the individual sciences images of each night are divided by their
corresponding masterflats.

\subsection{Bad pixels removal}

Before the combination of the full dataset of science images, it is necessary to identify those regions of each OSIRIS
CCDs where the quality of the image is degraded. To do that we have created a mask image (based on the normalized
masterflat) identifying: a) bad columns, b) hot pixels and c) vignetted regions of the camera (normally areas where the
count rate is less than 65\% of the peak). We have expanded slightly our masked region (to be conservative) to include
also the nearest pixels of those identified as bad ones. While doing the following reduction steps, our
individual science frames are masked with these masks.

\subsection{Astrometric calibration}

To avoid misalignments during the combination of our individual science images into a final mosaic, we need to assure
that the astrometry of all the individual images is the same. To conduct that task we use SCAMP
\citep{2006ASPC..351..112B}. SCAMP is used to put all our science images into a
common astrometric solution. SCAMP reads SExtractor catalogs and computes astrometric and photometric solutions for any
arbitrary sequence of FITS images in a completely automatic way. Our astrometric solution takes as a reference the
astrometry of the stars of the SDSS DR7 catalogue \citep{2009ApJS..182..543A} in our field of view. The number of stars
used in each science image for our astrometric solution is typically around a couple of dozens.

\subsection{Photometric calibration}

The photometric calibration of our science images is based on the photometry of SDSS DR7. We use the
(non-saturated) stars in the SDSS DR7 catalogue \citep{2009ApJS..182..543A} within our field of view. The
magnitudes of the stars in the SDSS DR7 catalogue that we have used are those from the PSF photometry. We
matched the SDSS DR7 photometric catalogue to ours, after which we multiplied our images making the photometry
in both catalogues equal. The multiplicative factor is chosen such that our images are calibrated to a common
zeropoint of 32 mag. The typical number of stars that are within each our individual science images to conduct
this photometric calibration task is $\sim$30.

\subsection{Sky determination}

The sky determination and subtraction is done for each of our science image individually before the final coaddition.
The determination of the sky is done using only those pixels of the images that are not identified as objects by
SExtractor. We place 10$^5$ apertures randomly located through the images and we determine the resistant mean value of
the counts in these apertures. We subtract that value to the calibrated images.

\subsection{Image coaddition}

Once the astrometry of every individual science image is recalculated to a common astrometric solution and the images
are calibrated as well as sky subtracted, we use SWarp \citep{2002ASPC..281..228B} to put all our data into a common
grid. SWarp is a program that resamples and co-adds together FITS images using any arbitrary astrometric projection
defined in the WCS standard. The combination uses the median of those images. The common field of view is illustrated in
Fig. \ref{weightmap}. The image resampling method that was used is LANCZOS3.

\begin{figure*}
\includegraphics[width=\textwidth]{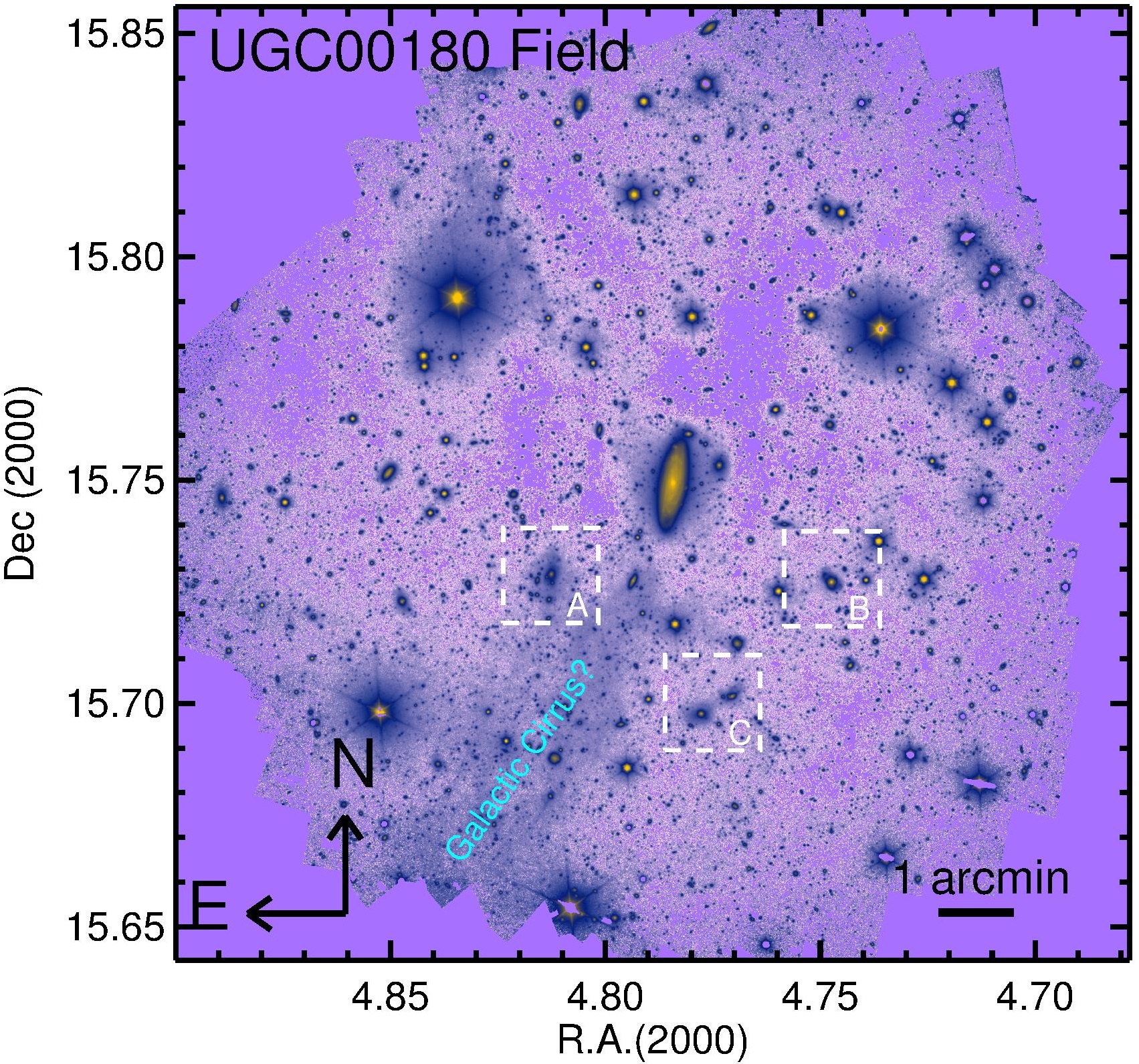}

\caption{Field of view of 12.7\arcmin$\times$12.7\arcmin around UGC00180. In addition to our main target, there are a
number of interesting astronomical objects that are highlighted. A zoom-in to objects tagged as a), b) and c) is shown in
Fig. \ref{zoomin}. The presence of an extended and filamentary emission in the bottom part of the image is also tentatively
identified as a Galactic Cirrus of our own Galaxy.}

\label{fieldofview}

\end{figure*}

The final coadded image (see Fig. \ref{fieldofview}) is significantly deeper than the individual exposures and low
surface brightness features, hidden in the individual exposures, emerge in the final stacking. These low surface
brightness features (extended dust emission, halos of bright stars, etc) affect the sky determination of our
individual science images. For this reason, it is necessary to mask these regions and repeat the process of the sky
determination in the individual exposures. The result of this repetition is our final image which we explore in detail
in the next sections.

\section{Results}
\label{results}

\begin{figure*}
\includegraphics[width=\textwidth]{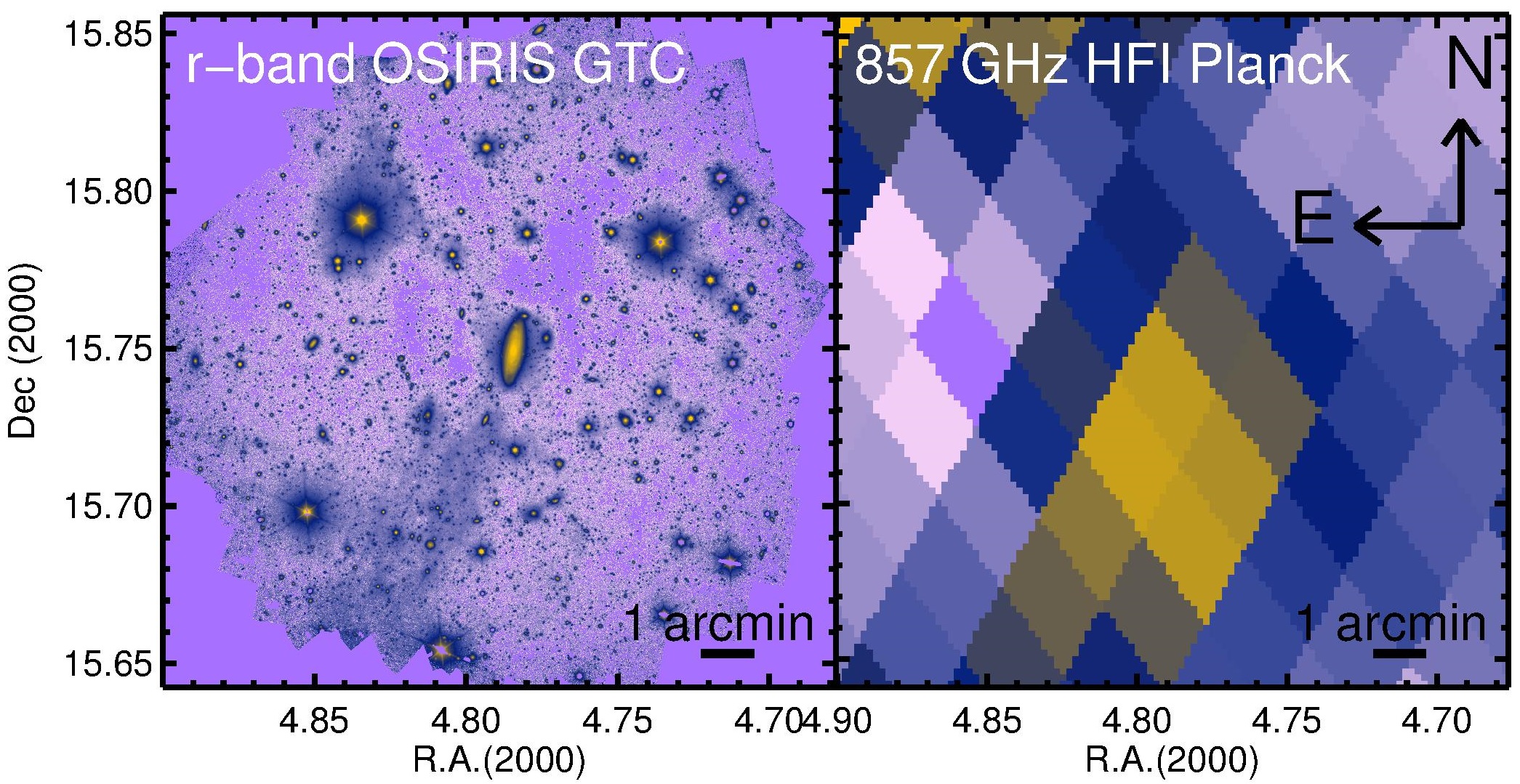}

\caption{The presence of Galactic Cirrus in our image. The figure compares the optical emission obtained with
GTC with the 350 $\mu$m map produced by the Planck satellite. The location of the maximum emission in the
Planck channel is similar to the position of the extended feature in the optical map. This suggests that the
origin of that extended emission could be produced by dust cirrus of our own Galaxy.}

\label{dust}

\end{figure*}

Our final image is shown in Fig. \ref{fieldofview}. Before discussing in detail the extended emission around
our main target, the nearby galaxy UGC00180, we shortly discuss here other visible structures in the field of
view around this object. The most conspicuous feature is the extended and filamentary emission observed in the
bottom part of the image that we have tentatively tagged as Galactic Cirrus. Without having color information, it
is complicated to identify the origin of this extended emission. We have used the Planck
satellite to see whether it is possible to see this feature at the 857 GHz (350 $\mu$m) channel
\cite[][]{2014A&A...571A...1P}. At this wavelength, the dust of our own Galaxy is particularly visible. The
spatial resolution in the far infrared image (1.7 arcmin/pixel) is so poor compared to the size of our image
that we can not make any clear statement on this. However, the position of the maximum emission of the dust
at the 857 GHz channel seems to coincide with the spatial location of the extended emission in the optical
(see Fig. \ref{dust}), suggesting a dust origin for this feature.

\begin{figure*}
\includegraphics[width=\textwidth]{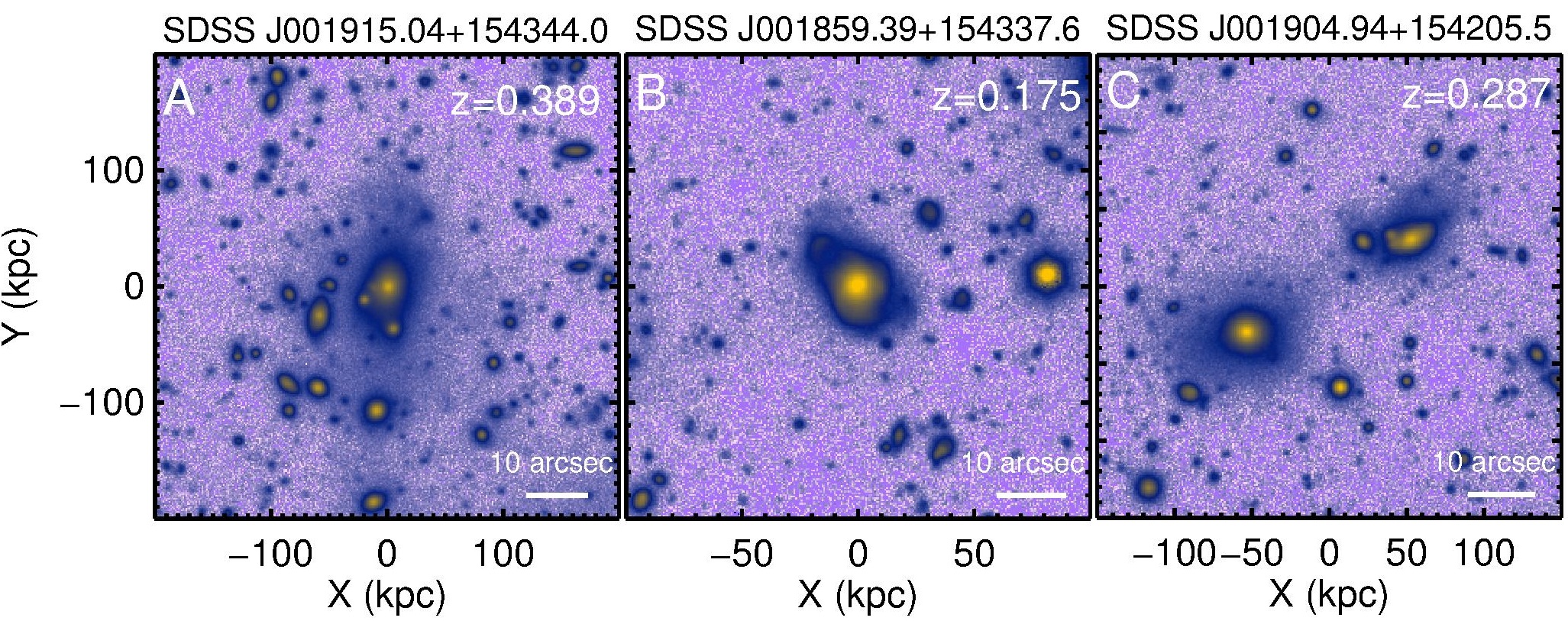}

\caption{Zoom-in to some of the interesting astronomical objects located nearby UGC00180. Panel A shows a cluster of
galaxy at z=0.389. The image is deep to enough to show the extended intracluster light of this object up to distances of
150 kpc. Panel B and C illustrate the merging activity in a galaxy located at z=0.175 and a galaxy pair at z=0.287. The
redshift of the examples shown in Panel A and B are spectroscopic. The
redshifts used in Panel C are photometric and obtained from the SDSS catalogue Photoz. More information can be found here
https://www.sdss3.org/dr10/algorithms/photo-z.php}

\label{zoomin}

\end{figure*}

We have also tagged three other regions in the image to illustrate the level of detail that can be explored in
our data. These regions are called A, B and C and have been shown in detail in Fig. \ref{zoomin}. Panel A shows a
galaxy cluster located at z=0.389. The region shown corresponds to 200$\times$200 kpc at the cluster redshift. The
presence of intracluster light (observed in the g'-band restframe) at distances as far as 150 kpc is very
remarkable (at this redshift the cosmological dimming is 1.43 mag) and illustrates the depth of this image. We will
quantify this depth in the following subsection. Panels B and C show galaxies undergoing mergers at different
redshifts. In the particular case of panel C, two galaxies of similar brightness at z=0.287 seem to have extended
stellar halos that are connected by a bridge of stars. The distance between these two galaxies is larger than 100 kpc.

\subsection{Surface brightness limit and comparison with other surveys}

To provide the limiting surface brightness of a given image, it is first necessary to define an area where a
given fluctuation is considered a detection or not. For instance, for the SDSS survey,
\citet[][]{2004AJ....127..704K}, using circular apertures of R=12\arcsec, found that a 3$\sigma$ fluctuation
in the surface brightness distribution of the image corresponds to an object with  $\mu_{lim}$=26.4
mag/arcsec$^2$ (g'-band). Alternatively, another way to explore the
limiting surface brightness of an image is to determine the limiting surface brightness down to which a
galaxy profile can be confidently explored. \citet[][]{2006A&A...454..759P}, using also SDSS, were able to extract
reliable (3$\sigma$) surface brightness profiles down to $\sim$27 mag/arcsec$^2$ at R=150\arcsec (g', r'
band). For the SDSS images, the number of pixels explored along a circular longitude of that radius is
equivalent to the number of pixels inside a circular aperture of R$\sim$11\arcsec. To put these numbers into
context, it is worth noting that this depth was obtained with a 2.5m telescope using $\sim$1 min exposure
time.

There have been increasing efforts in recent years to obtain very deep imaging of nearby galaxies. 
\cite{2015MNRAS.446..120D} have  summarized the depth of different projects and we refer to the reader to that
reference for an exhaustive summary of the current status in the literature. A large number of previous works,
including \cite{2015MNRAS.446..120D}, have been conducted using the Canada France Hawaii Telescope (CFHT).
Those works \citep[e.g.][]{2012ApJS..200....4F,2015MNRAS.446..120D} typically reach a depth of 28.5-29
mag/arcsec$^2$ (g' band) using 40-60 min in a 3.6m telescope. \cite{2010ApJ...709.1067B} using also the same
telescope but with integration of 5-10h claim detections of surface brightness features of $\sim$30
mag/arcsec$^2$ (g' band). Using the SDSS Stripe82 data obtained by the
SDSS 2.5m telescope during $\sim$1h, \citet{fliri2016} estimate a 3$\sigma$ detection (r' band) at 28.5 mag/arcsec$^2$.
\cite{2014ApJ...791...38W} claim a surface brightness limit of 29.5 mag/arcsec$^2$ (V band) with 10.25 hours on
source using the 0.6/0.9 CWRU Burrell Schmidt telescope. Finally, \cite{2010A&A...513A..78J} using the VLT
telescope during 6h exposure, obtained surface brightness profiles down to a limit of $\mu_R\sim30.6$ mag/arcsec$^2$ for
the nearby galaxy NGC3957.

Other works \citep[i.e.][]{2010AJ....140..962M}, using more modest apertures (D$<$0.5m) have reached 28-29 mag/arcsec$^2$ (V
band) using 10-15h on source. Finally, using an array of lenses equivalent to a 0.4m diameter telescope \citep[the Dragonfly
telescope;][]{2014PASP..126...55A}, \cite{2014ApJ...787L..37M} claim the detection of features with $\mu_g\sim29.5$
mag/arcsec$^2$ and $\mu_r\sim29.8$ mag/arcsec$^2$ on scales of $\sim$10 arcseconds for a total of 35 hours. 

\begin{figure*}
\includegraphics[width=\textwidth]{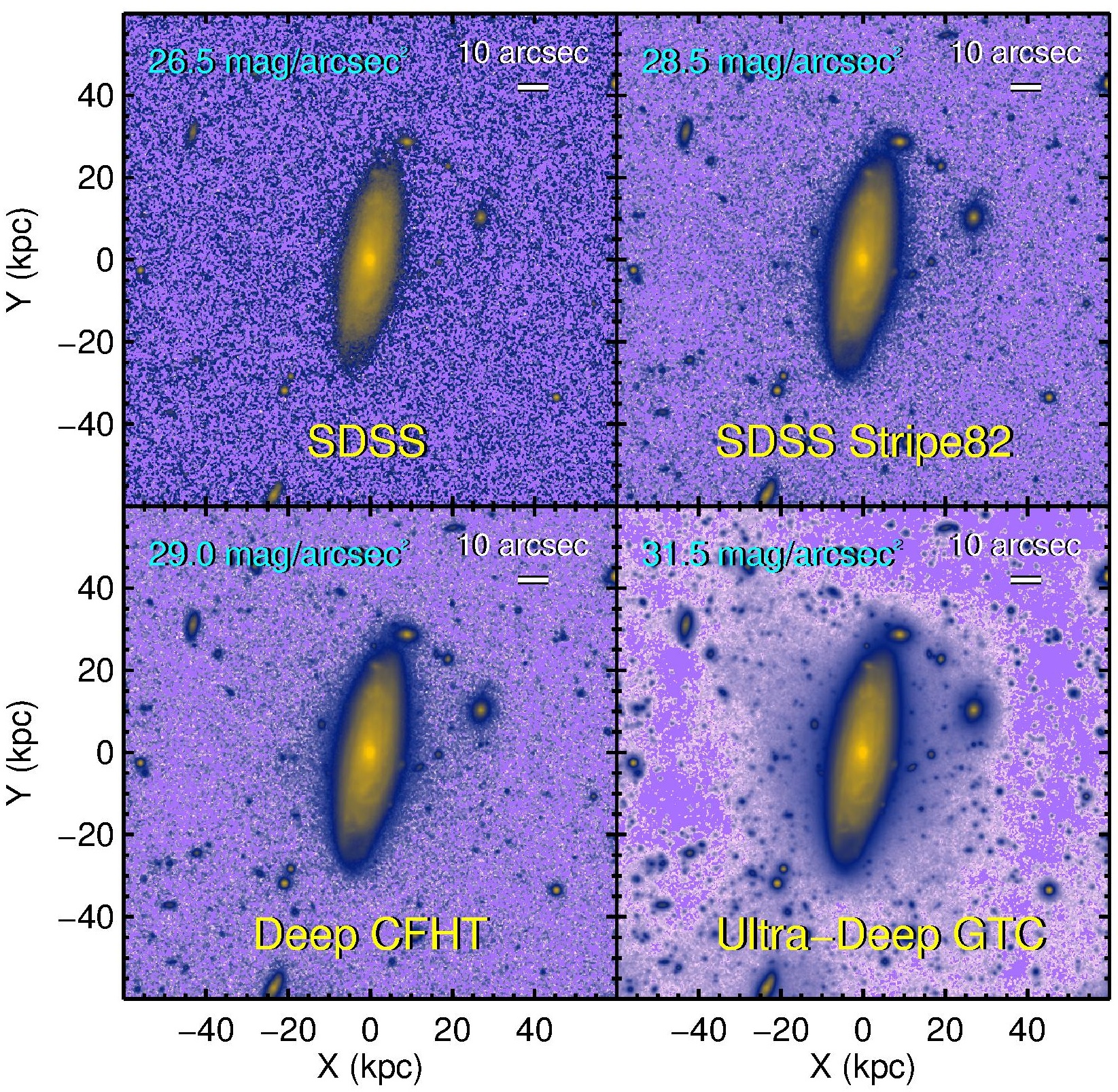}

\caption{UGC00180 as it would be observed by different surveys: SDSS, SDSS Stripe82, Deep data with CFHT
\citep[i.e.][]{2012ApJS..200....4F,2015MNRAS.446..120D} and the present work. Each surface brightness limiting magnitude has
been estimated as a 3$\sigma$ surface brightness fluctuation in boxes of 10\arcsec$\times$10\arcsec. Note how, for this
galaxy, the emergence of a stellar halo requires reaching limiting surface brightness fainter than 30 mag/arcsec$^2$ in the
r-band.}

\label{surfdepth}

\end{figure*}

In this work, we have decided to obtain the limiting surface brightness of the image as the equivalent
to a 3$\sigma$ fluctuation (compared to the sky noise) in square boxes of 10\arcsec$\times$10\arcsec.
The reason behind using boxes of this size is given by the typical size of the components we are
interested to explore in the stellar halo of UGC00180 (see Fig. \ref{surfdepth}). At the redshift of
UGC00180, this aperture is appropriate to probe features of 7.3$\times$7.3 kpc. Values like that are
typical of the FWHM of streams of nearby galaxies \citep[e.g.][]{2008ApJ...689..184M}. Using square
boxes of 10\arcsec$\times$10\arcsec \space  we obtain a surface brightness limit of 31.5 mag/arcsec$^2$
(3$\sigma$; r-band). This surface brightness limit is referred only to the innermost 4$\times$4 arcmin
of the image, where the effective amount of time on-source is 8.1h

In order to put our observations in comparison with other deep surveys, in Fig. \ref{surfdepth} we show
how our galaxy would be seen at the depth of SDSS, Stripe82 and deep CFHT surveys. To mimic the depth of
the different surveys, we have used our original GTC data and we have added noise to the image until we
get a limiting surface brightness depth as the one reported in the literature for the different surveys.
The limiting surface brightness (in the r-band) is estimated as a fluctuation of 3$\sigma$ using
10\arcsec$\times$10\arcsec \space boxes. We use the following limiting values: 26.5 mag/arcsec$^2$(SDSS), 28.5
mag/arcsec$^2$ (Stripe82) and 29 mag/arcsec$^2$ (Deep CFHT). We have checked that our  noise simulations
were conducted properly comparing our simulation of the SDSS depth with data of the same galaxy directly
obtained by the SDSS. Based on these noise tests, it is worth noting how, for UGC00180, the stellar halo
is only visible when the GTC depth is reached. 

\begin{figure*}
\includegraphics[width=\textwidth]{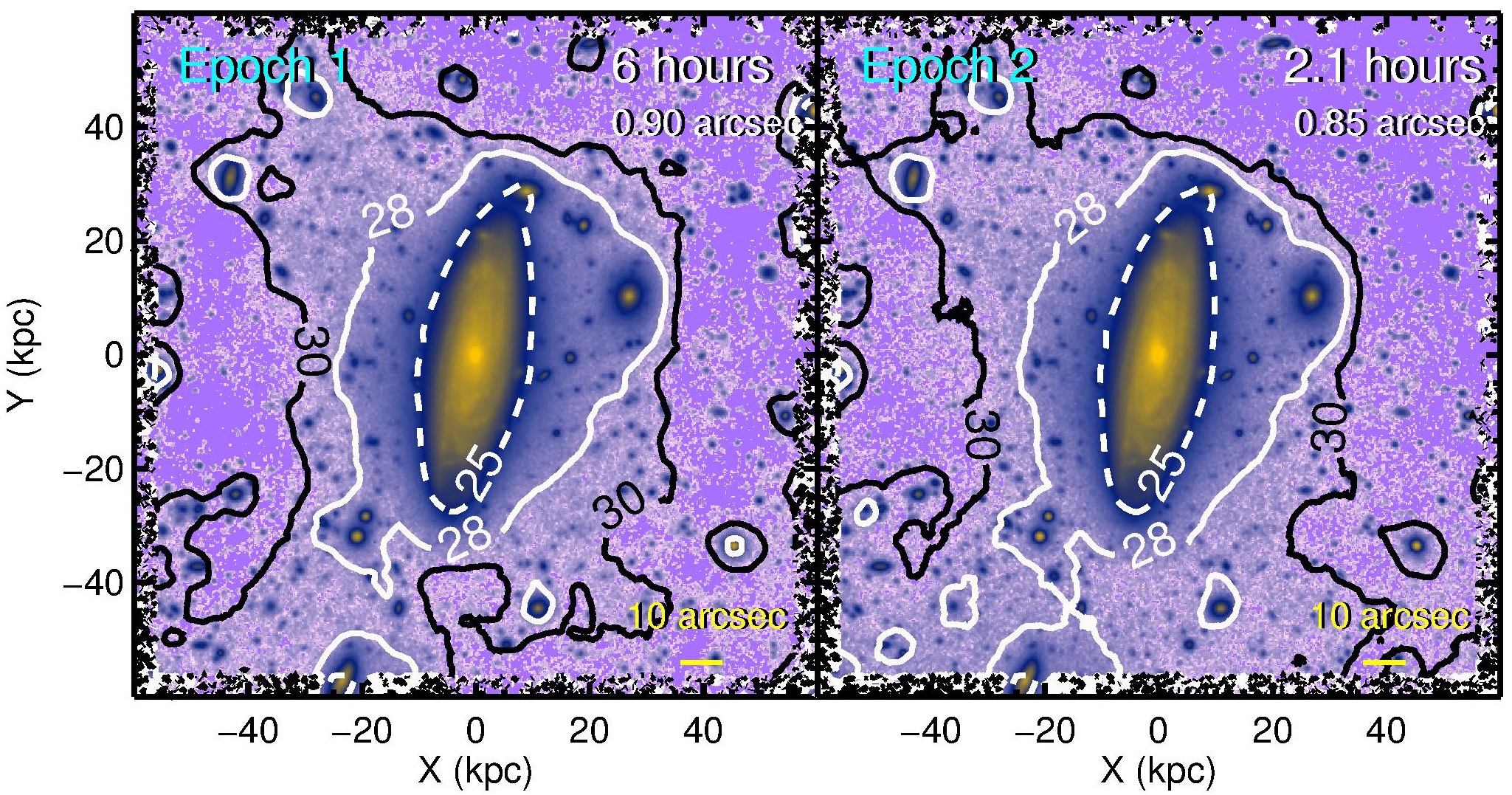}

\caption{This figure displays two independent observing datasets of the galaxy UGC00180 taken with GTC in
different dates. The figure shows
the repetability of the faintest features of the stellar halo of this galaxy. Overplotted in this
figure are the surface brightness contours corresponding to 25, 28 and 30 mag/arcsec$^2$.}

\label{repetability}

\end{figure*}

To further test the depth of our dataset, we explore the repetability of the faintest features we can
distinguish in our final image. The images of our galaxy were collected in two different observing sets with
different sky and seeing conditions. Consequently, those features that are visible in both blocks of
observations determine which is the actual surface brightness limit of our data in our shortest time exposure
set. The first set of data was taken during the 1st, 4th, 7th, 8th, 10th and 11th of November 2013 and it has a
total of 6 hours on source. The second set of data was collected during the 25th and 26th of November 2013 for a
total of 2.1 hours on source. A zoom-in around UGC00180 for the two sets of data is shown in Fig.
\ref{repetability}. Overplotted in this figure are the surface brightness contours corresponding to 25, 28 and
30 mag/arcsec$^2$. The vast majority of the faint features are identical in both independent datasets. There is
a small difference in the bottom right part of both images where the 30 mag/arcsec$^2$ extends a little bit
farther away in the shallow (2.1 hours) image with respect to the deeper block. This bottom part of the image
was slightly overexposed (in the shallow block) with respect to the rest of the image of the galaxy due to our
observational strategy. This produces a slightly higher S/N in this part of the image producing the observed
differences.

\section{Scattered light around UGC00180}
\label{scattered light}

As explained in Section \ref{data}, the reduction of the image has assumed a constant value for the sky. This value has been
determined after carefully masking all the individually detected objects and the low surface brightness extended features as
dust emission, halos of bright stars, etc. However, the scattered light produced by the convolution of the sources in the
field with the PSF constitute a complex background that needs to be explored to test whether the low surface brightness
features around UGC00180 have an origin external to the source. \cite{2009PASP..121.1267S} have conducted a detailed
analysis of the scattered light produced by the bright stars in a given field. They show that in fields like the center of
the Virgo Cluster (which contains three very bright stars of eighth magnitude to the west of M87) the convolution of the PSF
with the bright stars of the images implies that every single pixel of the image beyond 29 mag/arcsec$^2$ (V-band) is
dominated by the scattered light of one or more stars. If the contribution of scattered light is homogeneous over spatial
scales similar to the size of the object of interest, in our case UGC00180, then this scattered light is equivalent to have
a second sky level (the first one produced by the atmosphere at $\sim$22 mag/arcsec$^2$) overimposed over the galaxy.

\begin{figure*}
\includegraphics[width=\textwidth]{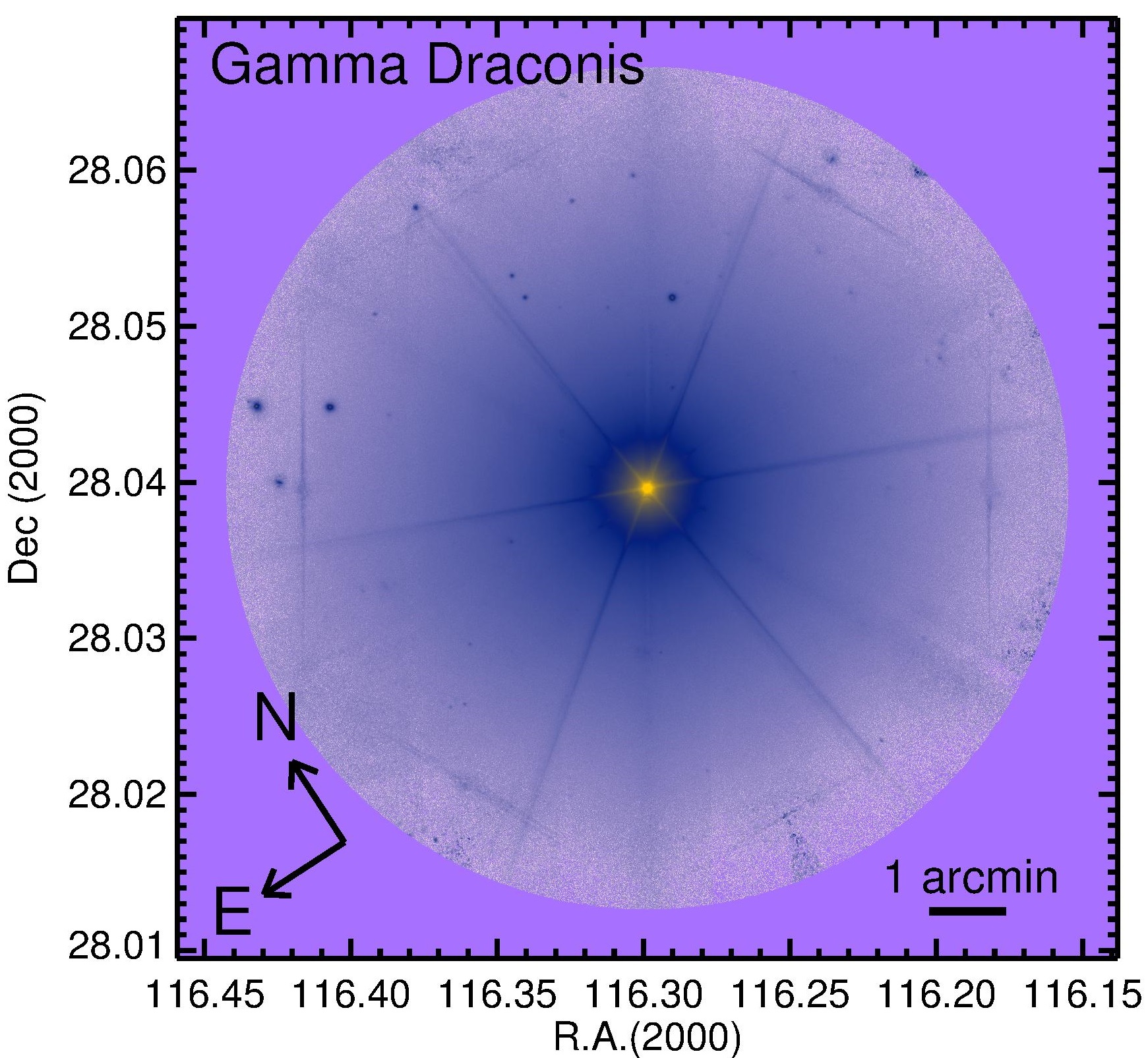}

\caption{The star $\gamma$ Dra (V=2.36 mag) as seen by the GTC telescope. This star has been used to model
the PSF of the GTC telescope in the r-band.}

\label{gammadraconis}

\end{figure*}

To construct the field of scattered light around UGC00180 produced by the bright stars we need to accurately characterize the PSF of the
image over an extension as large as possible. In practice, this means to characterize the PSF in this image at
least down to 5 arcmin which is the average separation of the brightest sources in our field. Ideally, one would
like to create such PSF using stars directly taken from the image and processed similarly. In practice, this is
extremely complicated as the presence of bright stars in the field is avoided to keep simple   the analysis
of the object under study. For this reason, we conducted a campaign for observing with GTC (using the same
rotation and dithering pattern than for the main object), the star $\gamma$ Dra (V=2.36 mag). The total amount
of time on source was 13.5 seconds (27 pointings of 0.5 seconds each using a dithering pattern of 9 pointings
and three position angles: 0, 120 and 240 degrees). Such a bright star allow us to explore the PSF\footnote{This
GTC PSF is available to the astronomical community at the following webpage:
$http://www.gtc.iac.es/instruments/osiris/osiris.php\#BroadBand_Imaging$. The data was collected on the 23th of
December 2014.} of the GTC telescope down to a radial distance of $\sim$5 arcmin. However, despite the very
short integration times we are using, the GTC PSF appears saturated in its innermost region ($<$10 arcsec). For
this reason, we combine that PSF with a PSF  built using the non-saturated PSFs of the UGC00180 field. Both PSFs
are matched to cover the entire brightness and radial range. The Fig. \ref{gammadraconis} shows  $\gamma$ Dra as
seen by the GTC telescope.

\begin{figure*}
\includegraphics[width=\textwidth]{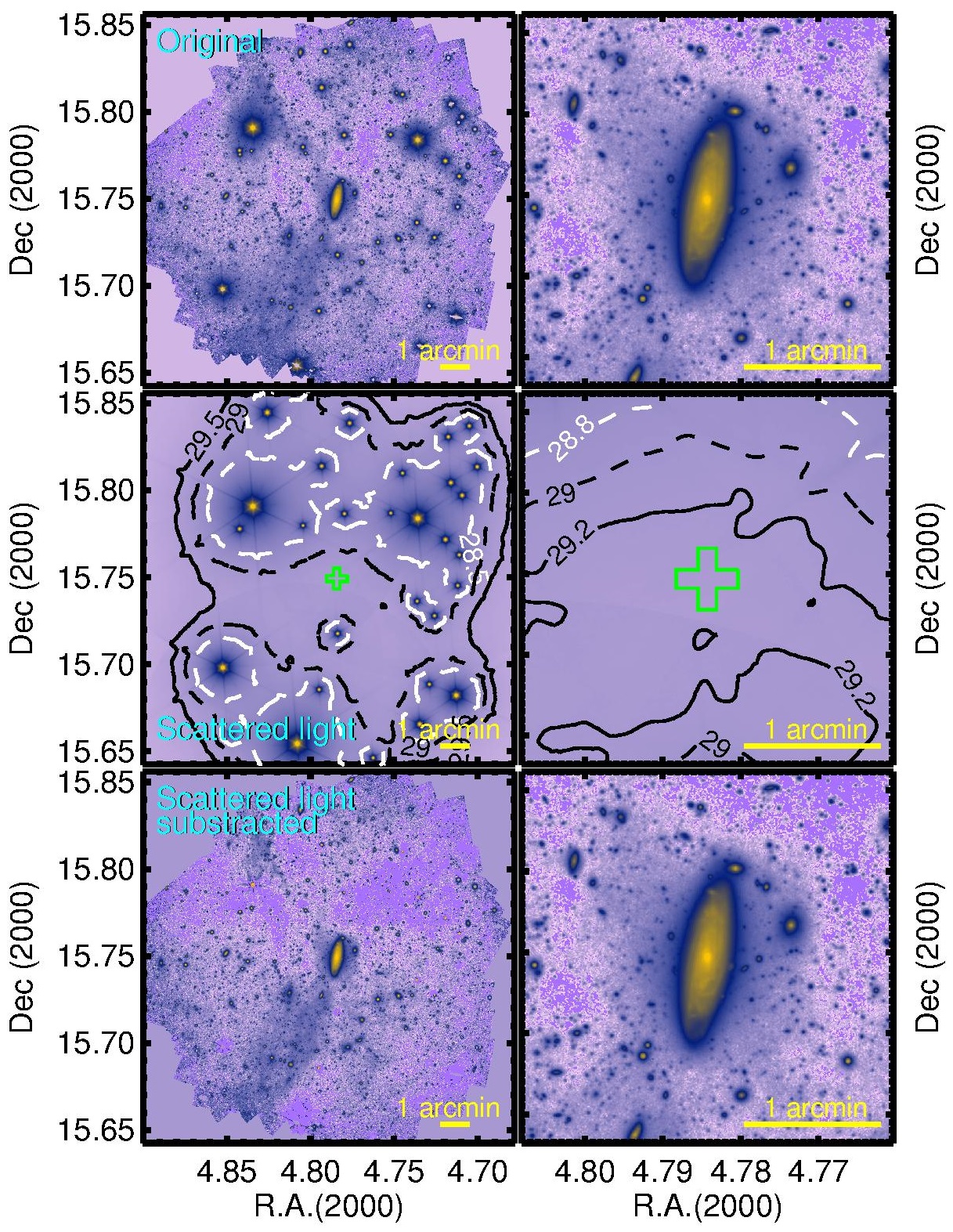}

\caption{The field of scattered light around the target galaxy UGC00180 produced by all the stars brighter than R=17 mag.
The original field and a zoom in to the galaxy is illustrated in the first row. The second row shows the field of scattered
light. The position of the galaxy is illustrated with a green cross. The contours of equal surface brightness 28.5, 29 and
29.5 mag/arcsec$^2$ (left column) and 28.8, 29 and 29.2 mag/arcsec$^2$ (right column) are shown. Finally, the third row is
the original field after the subtraction of the scattered light produced by the brightest sources. The shape of the stellar
halo of UGC00180 is unaffected by the scattered light around.}

\label{scatterfield}

\end{figure*}

With the PSF extremely well characterized  down to large radial distances (error less than 0.07\%, 0.11\% and 0.28\% at 1
arcmin, 3 arcmin and 5 arcmin respectively) we construct the field of scattered light produced by the brightest (R$<$17 mag)
stars of the image. It is worth noting that (except for UGC00180), all the sources of our image brighter than R=17 mag are
point-like sources. The selection of the bright stars in our field of view is done using the USNO
catalogue\footnote{http://www.nofs.navy.mil/data/FchPix/}. In particular, we have used the information provided by the  UCAC
3 catalogue \citep{2009yCat.1315....0Z} which contains the magnitude of the stars in R. We have used all the stars brighter
than the above magnitude within a radial distance of 7 arcmin to UGC00180. Once the catalogue of bright stars is
constructed, we build the scattered light field locating the GTC PSF (normalized to the flux provided by the USNO catalogue)
in each position where the bright stars are. The results of doing this is illustrated in Fig. \ref{scatterfield}. UGC00180
is placed in a region of the image where the contribution of the scattered light of the nearby brightest sources is rather
homogeneous and around 29.2 mag/arcsec$^2$. In fact, after subtracting the scattered light distribution (and adding back a
constant value to the sky to recover the zero value from the sky) there is not any effect on the structure of the stellar
halo around UGC00180. 

The field of scattered light created above is based under the assumption that the PSF shape
(particularly its wings) does not vary over the entire field. However, it is worth noting that the GTC
PSF was created to represent the effect of the PSF on the center of the image. A future study of the
stability of the GTC PSF mimicking the position of stars in different position of the final image would
be desirable to explore whether the above assumption is correct or not.

\section{The stellar halo of UGC00180}
\label{stellarhalo}

\begin{figure*}
\includegraphics[width=\textwidth]{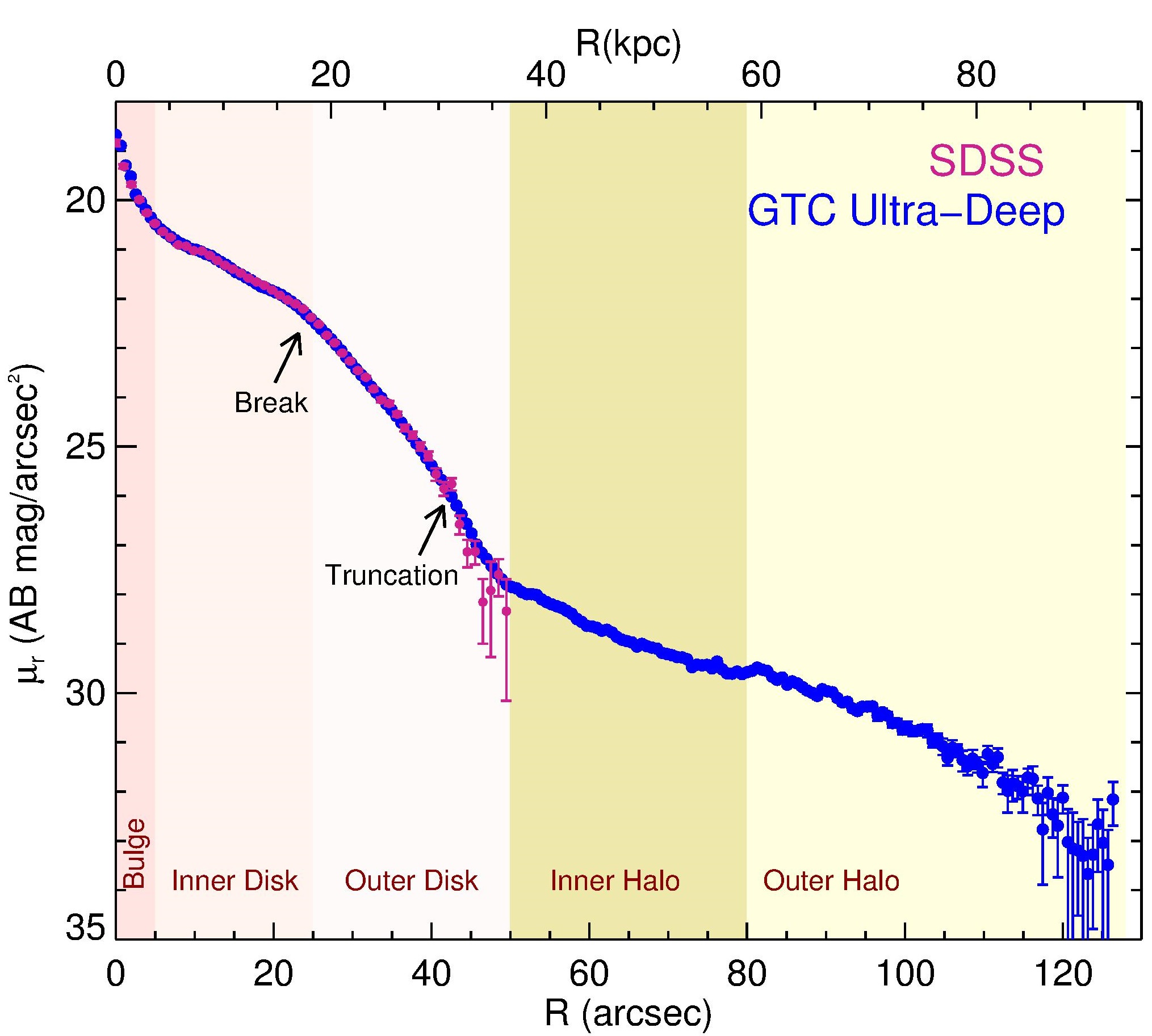}

\caption{The surface brightness profile of UGC00180 in r-band. The surface brightness profile of the galaxy
as seen by SDSS is overplotted (dark pink points). The depth of the GTC data allows to explore the surface
brightness profile of the galaxy 5 magnitudes deeper than with SDSS. The positions of the break of the inner
disk as well as some hint of an outer disk truncation is indicated with arrows.}

\label{profile}

\end{figure*}

The surface brightness profile of UGC00180 has been obtained through elliptical apertures with ellipticity
changing with radial distance to reflect better the disk and the outer (more roundish) component of the
galaxy. Nearby contaminant sources like foreground stars and background galaxies were identified with
SExtractor \citep{1996A&AS..117..393B} and masked. In those cases where SExtractor was unable to identify a
source, we masked the object manually. In addition, the Galactic Cirrus feature found in the south-east part of the
image was also masked to avoid its influence in the outer region of UGC00180. The resulting surface
brightness profile of UGC00180 is shown in Fig. \ref{profile}.

The depth of the GTC observations allow us to explore a range of $\sim$14 magnitudes in the galaxy profile, from
around 19 mag/arcsec$^2$ in the center down to $\sim$33 mag/arcsec$^2$ in the outskirts. We have also
overplotted the profile of the galaxy obtained using the SDSS image with the same elliptical apertures. The
GTC surface brightness profiles is 5 magnitudes deeper than the SDSS one. This is in excellent agreement with
the theoretical expectation taking into account the different in size of the telescopes (10.4m GTC vs 2.5m
SDSS) and the amount of time on source on the object (8.1h GTC vs 1 min SDSS). Note that the seeing in both datasets are
comparable ($\sim$ 1\arcsec).

We have divided our profile in 5 different spatial regions to illustrate the different structures of the galaxy. The
innermost region corresponds to the bulge of UGC00180. From 4 to 18 kpc we see a gentle exponential decline that we identify
with the inner disk of the object. After a break, the surface brightness profile continues declining exponentially down to
37 kpc (the outer disk). This kind of behavior (broken exponential) has been identified many times in the literature using
SDSS data \citep[see e.g.][]{2006A&A...454..759P}. In fact, the SDSS data of this galaxy also shows the break feature. Close
to the end of the outer disk we have a hint for a new down-bending "break" feature. We tentatively identify this as the
truncation of the disk. The faint surface brightness of the truncation ($>$26 mag/arcsec$^2$) makes it quite complicated to
identify in galaxies which are not completely edge-on using surveys like SDSS \citep[see a discussion about this
in][]{2014MNRAS.441.2809M}. The region beyond 37 kpc is dominated by a roundish component around the galaxy which is
declining exponentially. We call this region the stellar halo of the galaxy. There is a soft bump around 60 kpc and 30
mag/arcsec$^2$ which corresponds to the regions where the light distribution around the galaxy is more filamentary. In fact,
this feature can be identified with the tails of light with see in the north-east and south-west of the galaxy. The excess
of light in the south-east region is probably Galactic dust and, consequently, it has been masked during the analysis. We
use this radial position to identify two regions of the stellar halo: the inner and the outer part.

The stellar halo of UGC00180 is full of small clumps. These clumps could have potential different
origins: background galaxies, foreground faint stars, satellites of UGC00180 or even regions of star
formations. Not having color information at these extremely low surface brightness level we can not add
much to distinguish among the different scenarios. Future works, including more bands, should be able to
explore this issue and also  the conjecture raised in the Appendix by \citet[][]{2005ApJ...629..239B} where
they show that even at low metallicities (Fe/H=-1), there may be "gegenschein" from dust scatter.

\subsection{The effect of the PSF on the surface brightness profile}

There is growing evidence in the literature showing that the effect of the PSF can alter significantly the
amount of stellar light located in the periphery of the galaxies \citep[see
e.g.][]{2008MNRAS.388.1521D,2013MNRAS.431.1121T}. This effect has now been studied in detail by
\cite{2014A&A...567A..97S,2015A&A...577A.106S}. The results of these works indicate that the effect
of the PSF can mimic artificial stellar halos around the galaxies. Consequently, it is necessary to carefully
account for the PSF effects \citep{1983ApJS...52..465C} before any conclusion about the surface brightness distribution of our galaxy
UGC00180 is made.

To correct by the effect of the PSF, it is required to have a detailed description of the PSF of the image. In particular,
to be able to explore the effect of the PSF, it is mandatory to have an accurate characterization of the PSF to radial
distances, at least, as far as 1.5 times the radius of the galaxy \citep{2014A&A...567A..97S}. In our case, this means
having a PSF well described up to a radial distance of $\sim$3 arcmin. As we have explained before, we have the PSF of the
GTC telescope accurately characterized down to a radial distance of 5 arcmin. This is $\sim$10 times larger than the optical
size (R$_{25}$) of the galaxy.

\begin{figure*}
\includegraphics[width=\textwidth]{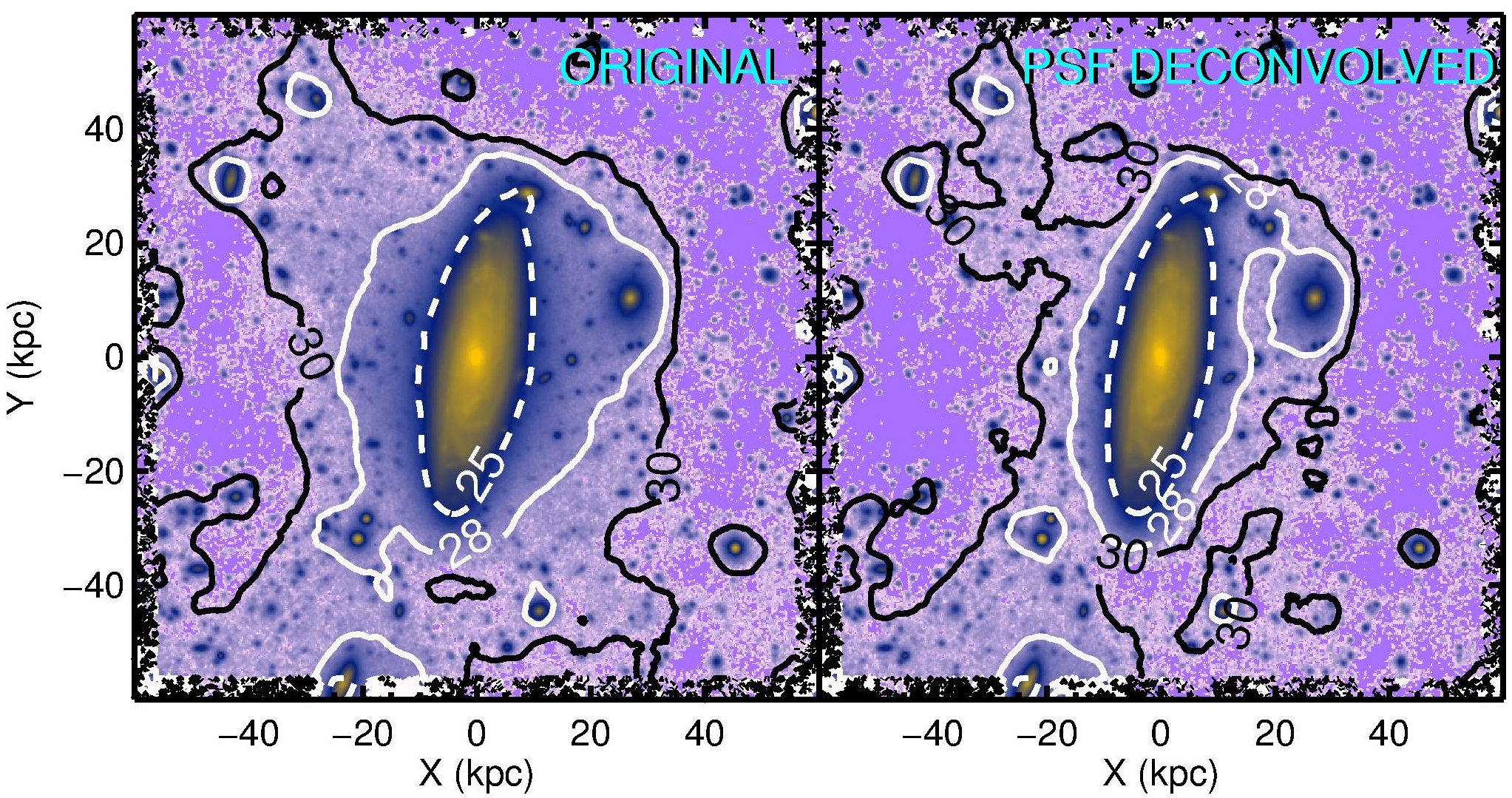}

\caption{The effect of the PSF on the surface brightness distribution of UGC00180. The figure shows the
dramatic effect of the PSF on the surface brightness distribution on the isophotal contours of the galaxy. We
indicate the position of the 25, 28 and 30 mag/arcsec$^2$ isophotes. The effect of the PSF
is particularly relevant when reaching surface brightness fainter than 25 mag/arcsec$^2$ (r-band).}

\label{contourpsf}

\end{figure*}

We simulate the effect of the PSF on the surface brightness distribution of our galaxy using the IMFIT code
\citep{2015ApJ...799..226E}. IMFIT is an image-fitting program specially designed for describing the surface
brightness distribution of the galaxies. We select the following functions to fit the light distribution of our
galaxy: a \cite{1968adga.book.....S} bulge, a broken disk exponential \citep{2008AJ....135...20E} and an
exponential stellar halo. We convolve these functions with the PSF of the image and fit the galaxy light
distribution. The convolved model is later on subtracted from the image to get the residuals of the fit. These
residuals include the spiral arm structure and other non-symmetric features which the model can not fit. After
that, we sum the residuals to the deconvolved IMFIT model to create an image of the galaxy with the effect of
the PSF removed. We show the difference between the original image and the deconvolved one in Fig.
\ref{contourpsf}.

\begin{figure*}
\includegraphics[width=\textwidth]{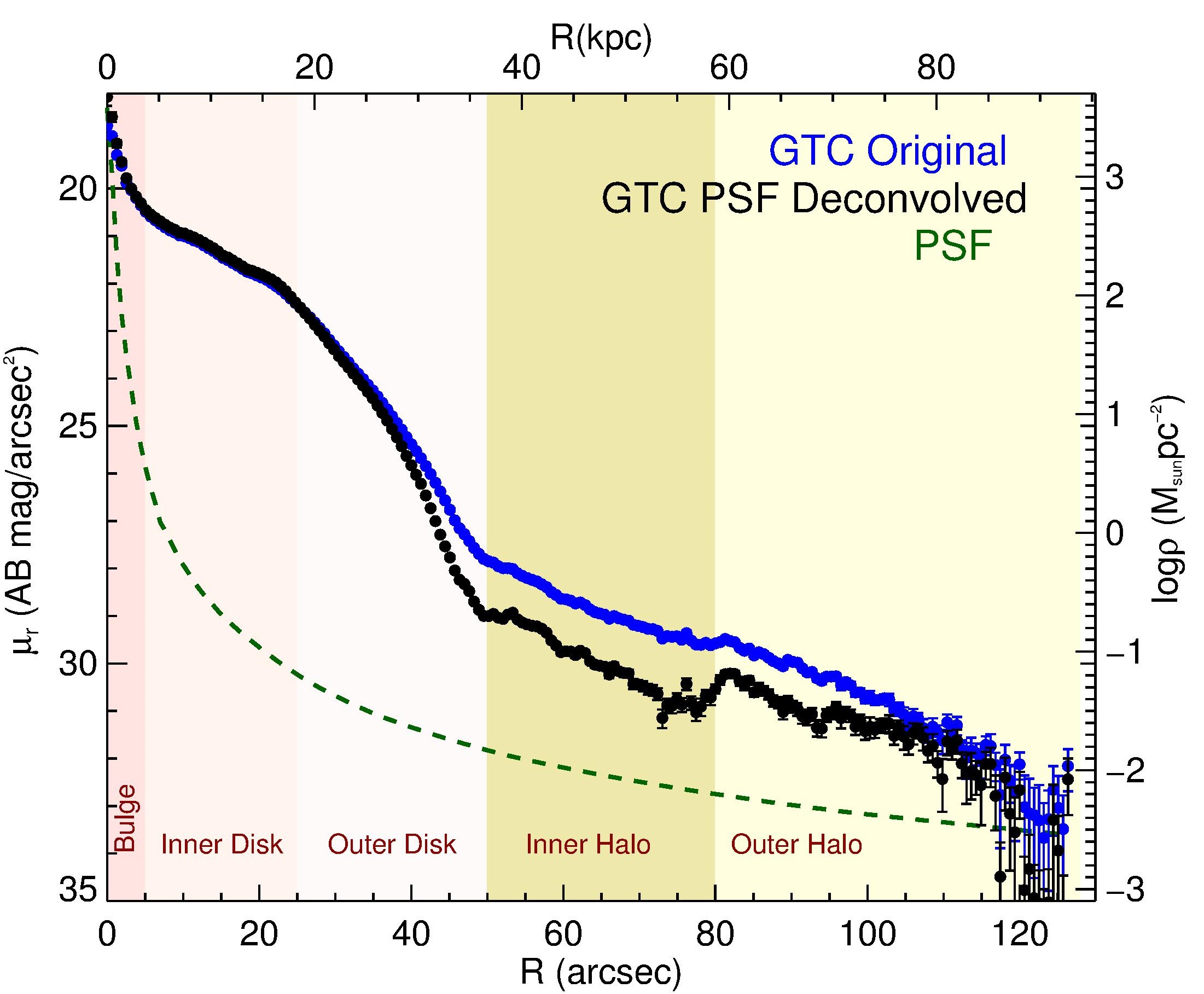}

\caption{The effect of the PSF on the surface brightness profile of UGC00180. The original profile is shown
using  blue points whereas the profile obtained after accounting by the effect of the PSF is plotted using
dark  points. The green dashed line shows the surface brightness profile of the GTC PSF.}

\label{profilepsf}

\end{figure*}

The effect of the PSF on the galaxy light distribution is particularly relevant beyond 25 mag/arcsec$^2$
(r-band). In fact,  Fig. \ref{contourpsf} shows that after correcting by the effect of the PSF, the 28
mag/arcsec$^2$  isophotal contour has a disk-like shape. This is strikingly different from the roundish shape
this contour has when exploring the original (i.e. PSF affected) image. The filamentary structure of the extra
light surrounding UGC00180 is also more evident once its light distribution is corrected by the effect of the
PSF. The effect of the PSF on the surface brightness profile of the galaxy is illustrated on Fig.
\ref{profilepsf}. The PSF affects both the central part, decreasing the central surface brightness of the bulge
by $\sim$1 mag/arcsec$^2$ as well as the very outer region of the galaxy where the effect is the opposite.
Beyond 25 mag/arcsec$^2$ the deconvolved profile starts to deviate from the original profile. At radial
distances further than 50 arcsec, the difference between the original and the PSF corrected profile is around
1  mag/arcsec$^2$. This has important consequences on the analysis of the galaxy: the stellar halo region is
much fainter (a factor of $\sim$2.5) than what it could be initially guessed using the original image of the
galaxy. We quantify this more precisely in the next section.

\subsection{The amount of stellar mass in the stellar halo}

Motivated by the shape of the light distribution in Fig. \ref{profilepsf}, we model the stellar halo light
distribution assuming an exponential profile. The use of an exponential law for describing the stellar halos
has been also followed in the past  \citep[e.g.][]{2005ApJ...628L.105I,2007ApJ...671.1591I}.
The use of an exponential model assumes that the amount of light in the stellar halo continues to rise towards the
center of the galaxy. In the central part of the galaxy, the light is dominated by the bulge and the disk,
outshining the contribution of the stellar halo. In this sense, the exponential behavior of the light
distribution of the stellar halo in the inner region is a hypothesis we have to assume here. However, this
growing contribution of the stellar halo towards the innermost region is motivated by numerical works exploring the
distribution of the accreted material in the galaxies \citep[see
e.g.][]{2010MNRAS.406..744C,2011MNRAS.416.2802F}.

We have measured the difference in the amount of light in the stellar halo one would have estimated using the
observed profile (i.e. the one affected by the PSF) and the amount of light in the deconvolved model. The results
of an exponential profile fitting to the halo light distribution are as follows. For the observed surface
brightness profile, affected by the PSF: central surface brightness $\mu_r(0)$=24.7$\pm$0.2 mag/arcsec$^2$ and
scale-length h=18.6$\pm$0.2 arcsec
(13.6$\pm$0.1 kpc). For the stellar halo corrected by the PSF: $\mu_r(0)$=26.6$\pm$0.3 mag/arcsec$^2$ and
h=23.5$\pm$0.2 arcsec (17.2$\pm$0.2 kpc). The stellar haloes estimated this way correspond to the following
fraction of light (r-band) compared to the total galaxy light: 0.11 (original profile) and 0.03 (stellar halo
corrected by the effect of the PSF). To allow a comparison of the stellar halo of UGC00180 with other stellar
halos reported in the literature, we assume that the ratio of the light contained in the stellar halo in the r-band is
similar to the ratio of stellar mass. This assumption is reasonable taking into account that the global color of
the galaxy (g-r)$\sim$0.8 is rather red (see Section 2), suggesting a low star formation activity, similar to what
one would expect in the stellar halo region.

\begin{figure*}
\includegraphics[width=\textwidth]{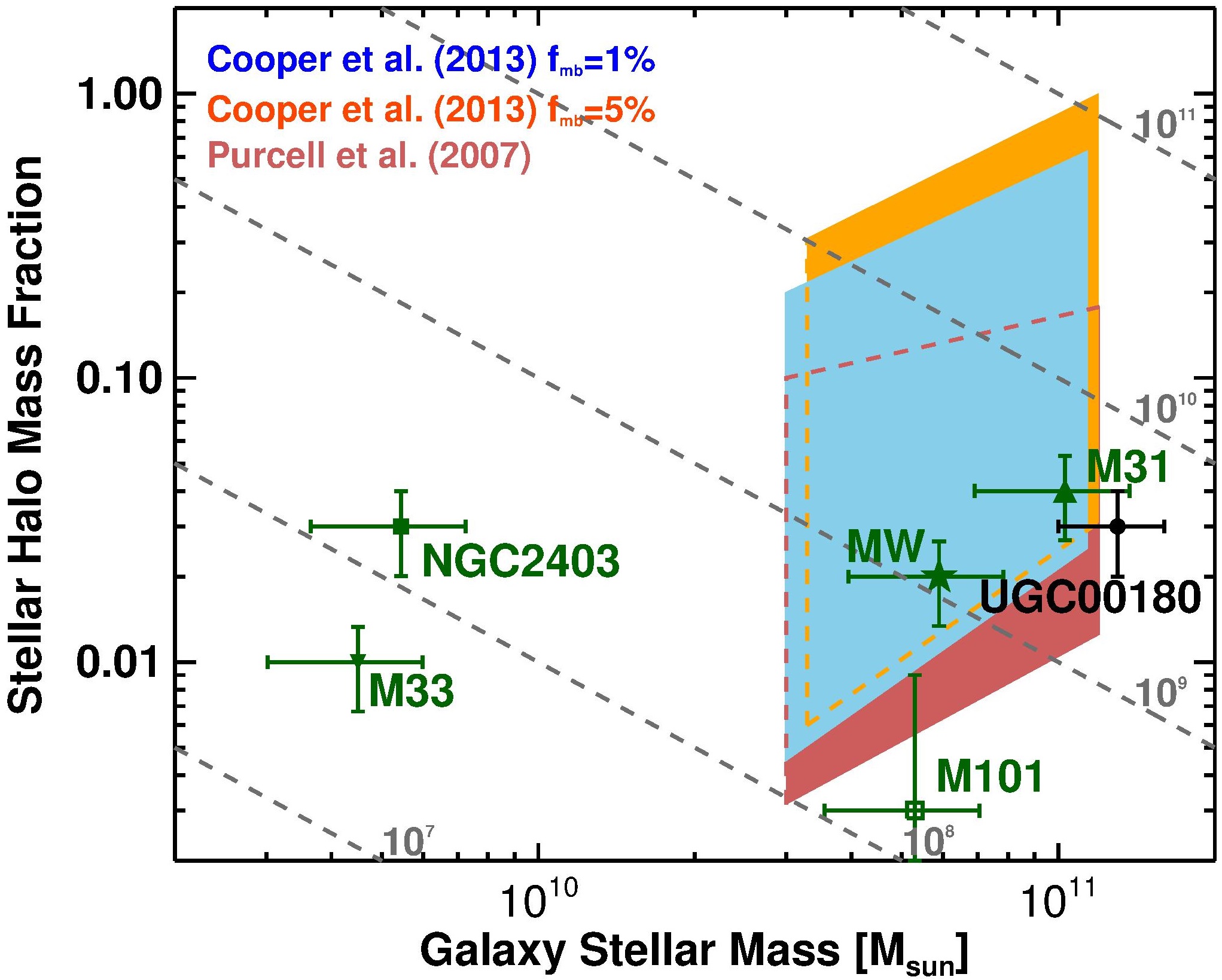}

\caption{Stellar halo mass fraction versus total galaxy stellar mass. The figure shows the location of the stellar halos
of galaxies compiled from the literature (green points) as well as the stellar halo of UGC00180 (black solid point). In
addition, we have overplotted the model predictions from \citet{2007ApJ...666...20P} (red area) and
\citet{2013MNRAS.434.3348C} (blue and orange) for galaxies inhabiting dark matter halos with
12$<\log_{10}M_{200}/M_{\sun}<$12.5. The grey dashed lines correspond to the positions on this plane of stellar
halos with fixed stellar mass (10$^7$, 10$^8$, 10$^9$, 10$^{10}$ and 10$^{11}$
M$_\odot$).}

\label{stellarhalos}

\end{figure*}

Figure \ref{stellarhalos} shows the stellar halo of UGC00180 in comparison with other stellar halos measured in
the literature: MW \citep[][]{2010ApJ...712..692C}, M31 \citep[][]{2011ApJ...739...20C}, M33
\citep[][]{2010ApJ...723.1038M}, NGC2403 \citep[][]{2012MNRAS.419.1489B} and M101
\citep[][]{2014ApJ...782L..24V}. The stellar halo mass fraction of UGC00180 is quite comparable to the one
measured in M31 (a galaxy with similar stellar mass). It is worth noting that this result is only achievable
when the effect of the PSF is taking into account. The amount of stellar mass contained in the stellar halo of
UGC00180 is around 4$\times$10$^9$ M$_{\sun}$. We will discuss this in the next section.

\section{Discussion}
\label{discussion}

Under the assumption of a Kroupa IMF and solar metallicity, the global color (g-r=0.78) of UGC00180 suggests an
average age for the stellar population of the galaxy of $\sim$8 Gyr \citep[][]{2010MNRAS.404.1639V}. This
implies a relatively quiet life for this galaxy and an assembly epoch at z$\gtrsim$1. It is theoretically
expected that the formation of stellar halos take place mostly at z$>$1
\citep[e.g.][]{2010MNRAS.406..744C,2011MNRAS.416.2802F,2012MNRAS.420..255T,2014MNRAS.439.3128T}. Observational
evidence for this has been reported by \citet[][]{2013MNRAS.431.1121T}. Assuming that the assembly history of
UGC00180 is typical for disk galaxies of its mass,  it is worth comparing the structure of its stellar halo
(size, shape, and amount of stellar mass) with the most recent cosmological simulations exploring this external
component.

\begin{figure*}
\includegraphics[width=\textwidth]{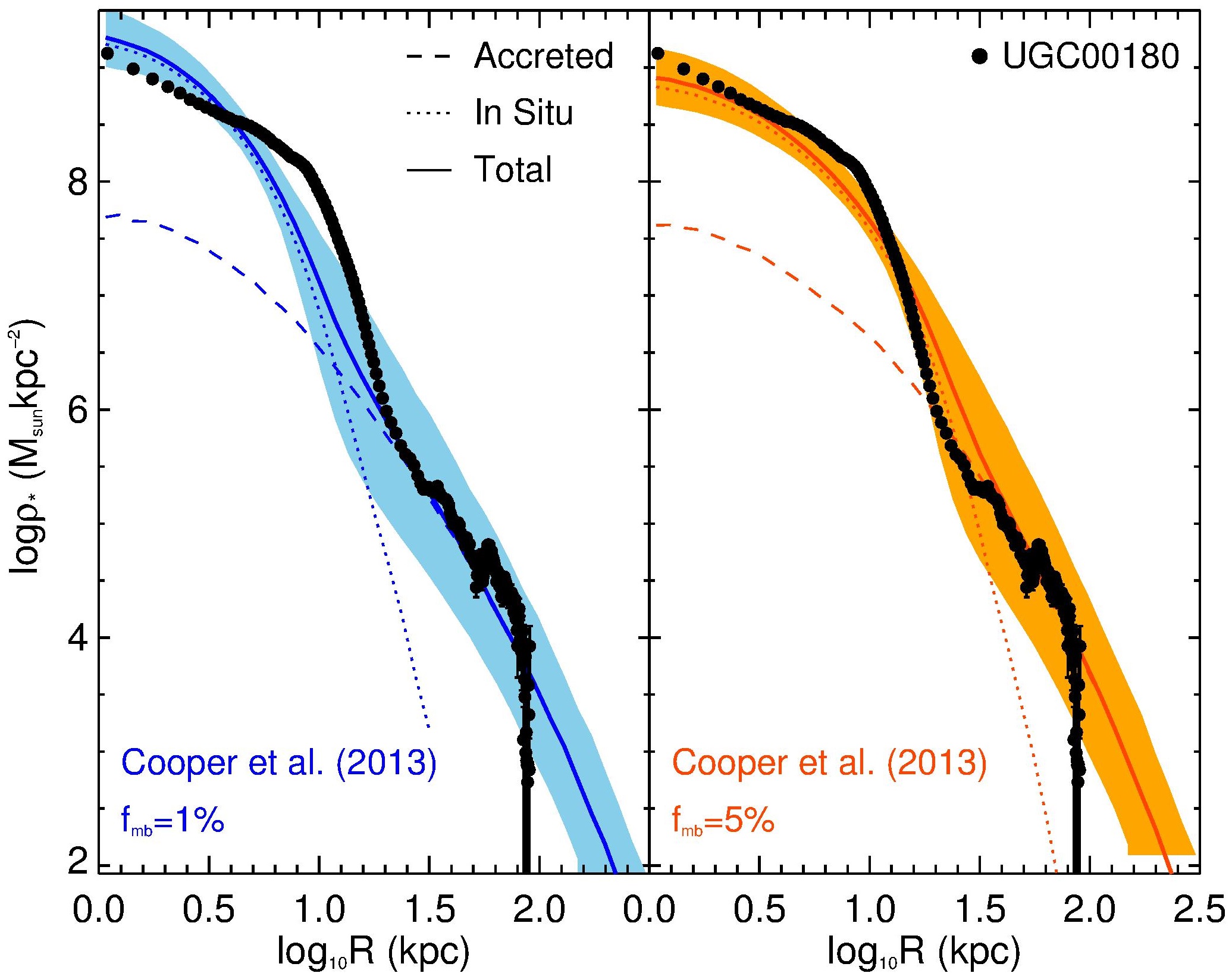}

\caption{Median profiles of circularly averaged stellar mass surface density, $\rho_*$, for accreted stars
(dashed lines) and in situ stars (dotted lines) taken from \citet[][]{2013MNRAS.434.3348C} model with
f$_{mb}$=1\% (left panel) and f$_{mb}$=5\% (right panel). A solid line shows the median profile combining
accreted and in situ components. The light blue and orange regions indicate the 10-90 percent scatter of the
median profiles. The black points correspond to the circularized (PSF deconvolved) stellar mass surface density
profile of UGC00180 assuming a constant M$_\star$/L$_r$=2.51 (see text for details).}

\label{coopermodel}

\end{figure*}

The comparison with the simulations is not straightforward as there is not a unique form of characterizing
which stars corresponds to the stellar halo and which ones to other components of the galaxy \citep[see
e.g.][]{2014MNRAS.444..237P}. In this paper we have used an exponential law \citep{1940BHarO.914....9P} motivated by the shape of the profile of the galaxy in
its outer region. In addition to the exponential shape,  other model profiles in the literature have been used
to fit the observed stellar halo distribution, such as a S\'ersic law, a Hernquist model or a power-law
\citep[see e.g.][]{2012ApJ...760...76G}. In Fig. \ref{coopermodel} we show our observed (circularized) profile
versus the results from the simulations by \citet[][]{2013MNRAS.434.3348C} for galaxies with dark halo virial
masses within the range 12$<$$\log_{10}$M$_{200}$/M$_{\sun}$$<$12.5. We have chosen this dark matter halo mass as
this is the one expected for galaxies like the MW and M31 \citep[e.g.][]{2010MNRAS.406..264W}. Taking into
account the stellar mass and morphological type of UGC00180, it is expected that this object also inhabit a
dark matter halo with similar properties.

Fig. \ref{coopermodel} shows two different renditions of \citet[][]{2013MNRAS.434.3348C} galaxy light
distributions depending on the "most-bound fraction" f$_{mb}$ used to model the location of stars within the
dark matter haloes. For instance, quoting \citet[][]{2013MNRAS.434.3348C}, f$_{mb}$=0.01 means that only the 1
per cent most-bound particles of the simulations are used to describe the stellar distribution. Comparing the
models with the light distribution of UGC00180, we can infer that the profile of the real galaxy fits rather
well with the expectation of the model with f$_{mb}$=0.05. The agreement is remarkably good (i.e. the observed
profile is within the scatter of the model prediction) down to $\log \rho_\star(M_\odot/kpc^2)\sim$3 or,
equivalently, R$\sim$80 kpc. Moreover, the models correctly predict an increasing relevance of the accreted
stellar component at radial distances beyond 25 kpc and $\log \rho_\star(M_\odot/kpc^2)<$5. This transition
region corresponds to an equivalent surface brightness of $\mu_r\sim$29 mag/arcsec$^2$. Beyond a radial
distance of 80 kpc ($\mu_r\gtrsim$32 mag/arcsec$^2$) the observed profile is quite uncertain, preventing us to
conclude whether the stellar halo of UGC00180 is "truncated" or whether the sudden drop of the stellar profile at
those distances is the result of an oversubtraction of the sky at such faint levels.

Independently of the shape of the profiles, it is worth comparing the model predictions about the amount of
stellar mass contained in the stellar halo of the galaxies. We have overplotted in Fig. \ref{stellarhalos} the
predictions from \citet[][]{2007ApJ...666...20P} (red area) and \citet[][]{2013MNRAS.434.3348C} (blue and
orange)  for galaxies inhabiting dark matter halos with 12$<$$\log_{10}$M$_{200}$/M$_{\sun}$$<$12.5. The
position of the models from \citet[][]{2013MNRAS.434.3348C} with f$_{mb}$=0.01 and f$_{mb}$=0.05 are indicated
respectively with blue and orange colors. Both the MW and M31 are in perfect agreement with all model
predictions. However, M33 and NGC2403 are significantly outside the theoretical regions. The reason why these
galaxies are not described by the models is easy to understand. We have only shown the model predictions for
dark matter halos with mass 12$<$$\log_{10}$M$_{200}$/M$_{\sun}$$<$12.5. However, M33 and NGC2403 inhabit dark
matter halos a factor of 10 less massive \citep[see e.g.][]{2011ISRAA2011E...4S}. The location of M101 is not
well described by \citet[][]{2013MNRAS.434.3348C} models, although it is statically compatible with
\citet[][]{2007ApJ...666...20P} predictions. In any case, this galaxy seems to have a very small stellar halo
for its total stellar mass. Could M101 be incorrectly described by the models due to its dark matter
halo mass is not within 12$<\log_{10}M_{200}/M_{\sun}<$12.5? According to Hyperleda, the maximum rotation
velocity corrected for inclination of M101 is V$_{rot}$=274$\pm$10 km/s. This is suggestive of a dark
matter halo mass similar to M31. However, taking into account that M101 seems to be clearly in
interaction, associating a large dark matter halo mass to its rotational velocity is, maybe, premature.
Finally, UGC00180 is slightly offset (although compatible within the error bars) from the predictions of the
theoretical models. The large stellar mass and rotational velocity of UGC00180 could reflect that this galaxy
inhabits a dark matter halo mass with $\log_{10}$M$_{200}$/M$_{\sun}$$>$12.5. If this were the case, the
agreement with  \citet[][]{2007ApJ...666...20P} and \citet[][]{2013MNRAS.434.3348C} models could be better.

According to Fig. \ref{stellarhalos}, MW, M31 and UGC00180 have stellar halos with mass ranging between 10$^9$ to
4$\times$10$^9$ M$_{\sun}$. Both \citet[][]{2007ApJ...666...20P} and \citet[][]{2013MNRAS.434.3348C} models predict that
the progenitors of the stellar halos of these galaxies will be satellites with M$_\star\sim$3$\times$10$^8$ M$_{\sun}$.
Consequently, on average, we expect a number of merging events with these types of satellites ranging from $\sim$3 (for
galaxies like MW) to $\sim$12 (for M31 and UGC00180). Nowadays, the number of satellite galaxies that both the MW and M31
have with M$_\star\sim$3$\times$10$^8$ M$_{\sun}$ and   within R$<$300 kpc is $\sim$1 \citep{2012AJ....144....4M}. If this
number is representative of the population of satellites of these galaxies in the past, we can infer that the average
merging timescale for these satellites is $\sim$1-3 Gyr (assuming MW and M31 are basically formed since z$\sim$2).

Finally, we can focus our attention on the general prediction by galaxy formation models stating that all
present-day galaxies will show several streams and a prominent stellar halo if they are observed down to
$\mu_V$$>$31 mag/arcsec$^2$ \citep[e.g.][]{2010MNRAS.406..744C}. In  MW, M31 and
UGC00180, where the observations have gone deep enough to explore this prediction, this is indeed the case. This
question remains open for M101, a galaxy with similar stellar mass to the ones above, whereas the deepest current
observation \citep[e.g.][]{2013ApJ...762...82M,2014ApJ...782L..24V} are at the limit ($\sim$30 mag/arcsec$^2$;
3$\sigma$ 10$\times$10 arcsec boxes) to explore this issue.

\section{Summary and Conclusions}
\label{conclusions}

In this paper we have addressed the following question: what is the surface brightness limit that current largest optical
telescopes ($\sim$10 m) can achieve within a reasonable amount on time (i.e. $<$10 hour on source)? Using the 10.4m GTC
telescope, during a total amount of time of 8.1 hours on source, we have found that is feasible to reach 31.5 mag/arcsec$^2$
(3$\sigma$; 10$\times$10 arcsec boxes in the r-band). This is a surface brightness limit around 1.5-2 mag deeper than most
current surveys dedicated to explore the faintest astronomical extended structures using integrated photometry.

Using this ultra-deep observation, we have explored the stellar halo of UGC00180, a galaxy with similar mass and
morphology to M31. After addressing the effect of the PSF on the surface brightness distribution of
this galaxy, we have been able to probe the surface brightness profile of this object down to
$\mu_r$$\sim$33 mag/arcsec$^2$. This is equivalent to the depth reached when using star counting technique to measure the
light profiles of galaxies, but this time for a galaxy located at 150 Mpc, where this technique is unfeasible.
 The fraction of light contained in the stellar halo of UGC00180 is
3$\pm$1\%, in agreement with state-of-the-art galaxy formation models. Our pilot project shows that current technology will allow us to study
the stellar halos of many hundreds of galaxies. This opens the possibility to explore the expected large
variety of shapes and morphologies for this faint component around galaxies. Reproducing, quantitatively, the
characteristics of the stellar halos in a large number of objects will be one of the most
demanding test for the $\Lambda$CDM galaxy formation scenario in the future.



\acknowledgments

We thank the referee for his/her constructive comments. Based on observations made with the Gran
Telescopio Canarias (GTC), installed at the Spanish Observatorio del Roque de los Muchachos of the
Instituto de Astrofísica de Canarias, in the island of La Palma. This work has been supported by the
``Programa Nacional de Astronom\'{\i}a y Astrof\'{\i}sica'' of the Spanish Ministry of Science and
Innovation under grant AYA2013-48226-C3-1-P.  We are grateful to Ricardo Genova for helping us with the
Planck data and Lee Kelvin for interesting input on the analysis of the images. Massimo Capaccioli,
Christer Sandin and Jovan Veljanoski did several stimulating comments on different aspects of this work.
We thank also Andrew Cooper by providing his data from his simulations to allow a direct comparison with
the observed profile. We are indebted with the support astronomers of the GTC telescope, in particular
with Antonio Cabrera Lavers for his wonderful job on getting this dataset.

\end{document}